\documentclass[twocolumn,amsmath,superscriptaddress,amssymb,amsfonts,aps,floatfix]{revtex4}

\usepackage{graphicx,color,rotating} 
\usepackage{bm} 

\begin{document}
  \title{Tunneling and Relaxation of Single Quasiparticles in a Normal-Superconductor-Normal Single Electron Transistor}

\author{Andreas~Heimes}
\affiliation{Institut f\"ur Theoretische Festk\"orperphysik and DFG-Center for Functional Nanostructures, Karlsruhe Institute of Technology, D-76128 Karlsruhe, Germany}

\author{Ville~F.~Maisi}
\affiliation{Low Temperature Laboratory (OVLL), Aalto University School of Science, P.O.~Box 13500, 00076 Aalto, Finland}
\affiliation{Centre for Metrology and Accreditation (MIKES), P.O. Box 9, 02151 Espoo, Finland}

\author{Dmitri~S.~Golubev}
\affiliation{Low Temperature Laboratory (OVLL), Aalto University School of Science, P.O.~Box 13500, 00076 Aalto, Finland}
\affiliation{Institute of Nanotechnology, Karlsruhe Institute of Technology, D-76021 Karlsruhe, Germany}

\author{Michael~Marthaler}
\affiliation{Institut f\"ur Theoretische Festk\"orperphysik and DFG-Center for Functional Nanostructures, Karlsruhe Institute of Technology, D-76128 Karlsruhe, Germany}

\author{Gerd Sch\"on}
\affiliation{Institut f\"ur Theoretische Festk\"orperphysik and DFG-Center for Functional Nanostructures, Karlsruhe Institute of Technology, D-76128 Karlsruhe, Germany}
\affiliation{Institute of Nanotechnology, Karlsruhe Institute of Technology, D-76021 Karlsruhe, Germany}

\author{Jukka P. Pekola}
\affiliation{Low Temperature Laboratory (OVLL), Aalto University School of Science, P.O.~Box 13500, 00076 Aalto, Finland}

\begin{abstract}
We investigate the properties of a hybrid single electron transistor, involving a small superconducting island sandwiched between normal metal leads, which is driven by dc plus ac voltages.
In order to describe its properties we derive from the microscopic theory a set of coupled equations. They consist of a master equation for the probability to find excess charges on the island, with rates depending on the  distribution of non-equilibrium quasiparticles. Their dynamics follows from a kinetic equation which accounts for the excitation by single-electron tunneling as well as the relaxation and eventual recombination due to the interaction with phonons. Our low-temperature results compare well with recent experimental findings obtained for ac-driven hybrid single-electron turnstiles.

\end{abstract}

 \maketitle

The excitation of non-equilibrium quasiparticles in superconductors of reduced dimensions
by an applied dc bias or ac radiation has been the subject of theoretical and experimental studies for decades.
It has been demonstrated, e.g., that quasiparticles excited by strong ac radiation may enhance both the 
critical current of superconducting bridges \cite{Wyatt} and the value of the superconducting gap 
\cite{Kommers,ESSS,Klapwijik}.
It has also been shown that a dc bias voltage applied to a metallic dot coupled to superconductors may 
lead to electronic cooling \cite{Refrigerator1,Refrigerator2}.
More recently, the issue of non-equilibrium quasiparticles has drawn renewed attention.
On one hand, it turned out that they reduce the coherence time of superconducting qubits 
\cite{cite:Qubit_experimental,cite:Qubit_theoretical}. On the other hand, they limit
the accuracy of single-electron turnstiles when they are used as current standards
\cite{Pekola_2008,Knowles_2012,Maisi_2009}. 

\begin{figure}[!ht]
	\begin{center}
\begin{tabular}{c}
\includegraphics[width=\columnwidth]{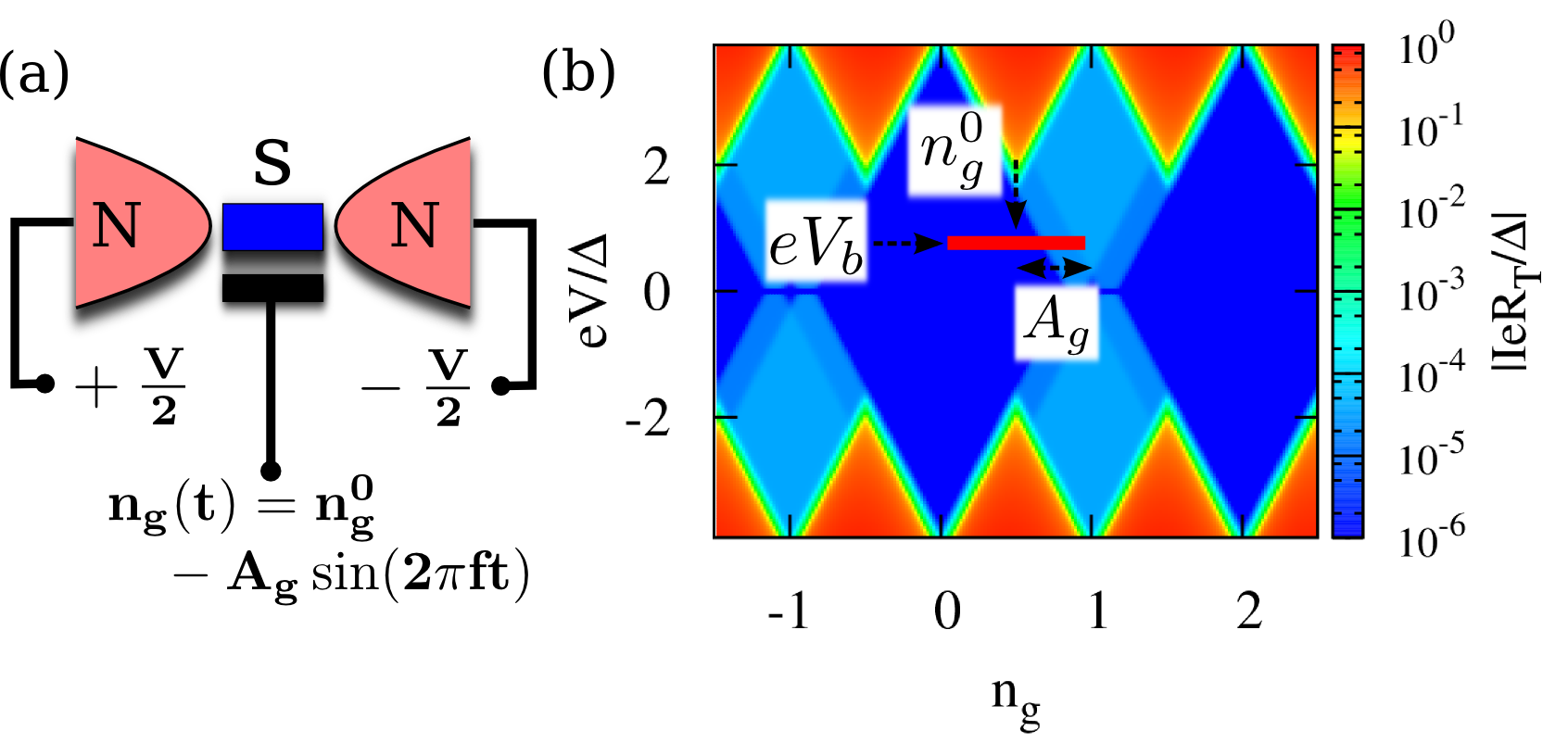} 
\\
\includegraphics[width=4cm]{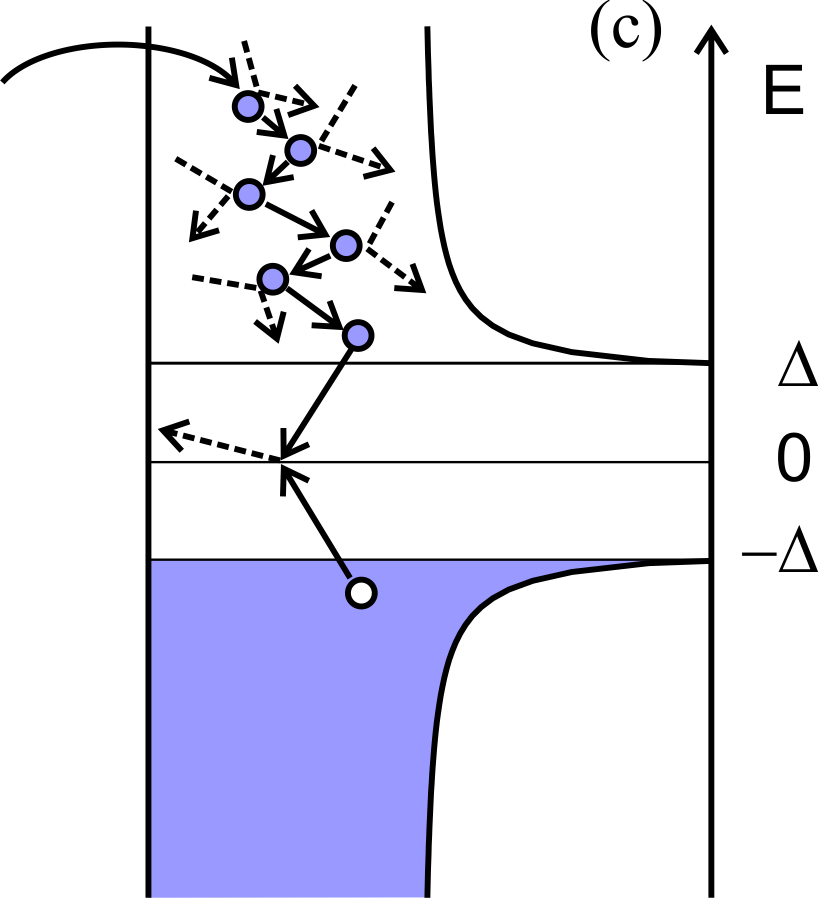}
\end{tabular}
\caption{(a) Schematics of a NSN single electron transistor. 
(b) Stability diagram in equilibrium: shown is the absolute value of the dc-current $I$ in units of $\Delta/eR_T$ where $\Delta$ is the superconducting gap, $e$ the electron charge and $R_T$ the tunneling resistance. During the turnstile operation a bias voltage $eV_b$, a dc gate offset $n_g^0$ between charging states $0$ and $1$ and an ac gate-modulation with amplitude $A_g$ are applied. (c) Illustration of the processes taken into account in our model: an electron-like quasiparticle
is injected into the island through one of the junctions, it is then scattered by phonons quickly relaxing to an energy just above $\Delta$,
and, finally, it recombines with a hole.}
\label{fig::setup_stability_diag}
\end{center}
\end{figure}

Experiments with qubits and turnstiles are usually performed at low temperatures and
 bias voltages, with superconducting grains of small size. Under these conditions
the number of excited non-equilibrium quasiparticles is low. Moreover, it is possible
to detect even a single quasiparticle trapped in a superconducting grain \cite{Maisi_2012}. 
In this limit the quasiclassical theory of non-equilibrium superconductivity based, e.g., on the Eilenberger or Usadel equations \cite{quasiclassical} is not sufficient. In this paper we extend this theory, starting from the microscopic theory of superconductivity but including the effect of 
single-electron charges and Coulomb blockade. Specifically, we consider a normal 
metal--superconductor--normal metal (NSN) single-electron transistor (SET) as depicted in Fig.~\ref{fig::setup_stability_diag}(a).
This setup has been used in  recent single-electron pumping experiments \cite{Maisi_2012}, and one of the
goals of our paper is to analyze them quantitatively.
As illustrated in the stability diagram in Fig.~\ref{fig::setup_stability_diag}(b) we assume that 
the SET is biased with a small dc voltage, 
and at the same time a sinusoidal ac drive is applied to its gate electrode. 
We derive a system of coupled equations which describe both the electron tunneling
into and out of the superconducting dot, the excitation of non-equilibrium quasiparticles and their relaxation and recombination due to inelastic scattering with phonons, see Fig.~\ref{fig::setup_stability_diag}(c). 

The paper is organized as follows.
In Sec. \ref{section::Model} we describe our model and derive a set
of equations, a master equation for the probability of finding excess charges on the island, coupled to a kinetic equation for the non-equilibrium quasiparticle distribution. Together they fully describe the non-equilibrium NSN SET.
We solved these equations numerically with results to be
presented in Sec. \ref{sec::results}. 
We also derive approximate descriptions and recover several results obtained earlier.   
In Sec. \ref{sec::summary} we will summarize our results. Some
details of the calculations are moved to the appendix.

\section{Model}
\label{section::Model}

We consider the system depicted in Fig.~\ref{fig::setup_stability_diag} (a), 
consisting of a superconducting quantum dot ($D$) coupled to the left 
($L$) and the right ($R$) bulk normal leads via tunnel junctions.
It is described by the Hamiltonian 
\begin{eqnarray}
 H=\sum_{r=L,\,R}H_r + H_D + H_T.
 \label{Ham}
\end{eqnarray}
The normal metal leads are assumed to be reservoirs of noninteracting electrons,
\begin{align}
 H_r=\sum_{k\sigma} \,(\xi_{rk\sigma}-\mu_r)\, c_{rk\sigma}^\dagger c_{rk\sigma}.
\end{align}
Here
$\xi_{rk\sigma}$ is the energy of an electron with momentum $k$ and spin $\sigma$, and 
$c_{rk\sigma}^\dagger$ are the corresponding electron creation operators.
The applied voltage shifts the electro-chemical potentials by $\mu_r=\pm eV/2$.

The Hamiltonian of the superconducting island accounts for the superconductivity,
the Coulomb interaction between electrons, and the electron-phonon interaction. It reads
\begin{align}
 H_D&=\sum_{k\sigma} E_k \,\gamma^\dagger_{k\sigma}\gamma_{k\sigma}  + E_C (\hat n-n_g)^2 \nonumber \\ &+ \sum_q \omega_q b_q^\dagger b_q + H_{e-ph}. 
\label{HD}
\end{align}
Here $E_{k}=\sqrt{\Delta^2+\xi_k^2}$ is the quasiparticle energy, 
$\Delta$ is superconducting gap, $\xi_k$ are the electron energies in the normal state,
while $\gamma^\dagger_{k\sigma}$ and $\gamma_{k\sigma}$ are the creation and annihilation operators of the
quasiparticles obtained after the Bogoliubov transformation known from the BCS theory. 
The second term of Eq. (\ref{HD}) describes the Coulomb interaction 
modeled by an effective capacitance and charging energy $E_C$. It depends on 
the number of excess electrons in the dot, given by the operator $\hat n$, and the dimensionless offset charge $n_g=C_gV_g/e$ induced by the gate voltage $V_g$ applied to the dot via the gate capacitance $C_g$.
The third term is the Hamiltonian of free phonons with frequencies  
$\omega_q$ and momenta $q$. Here and below we set $\hbar=k_B=1$.
Finally,  $H_{e-ph}$ describes the electron-phonon interaction.
After the Bogoliubov transformation to the quasiparticles it can be written in the form
\begin{align}
 H_{e-ph} = & \sum_{qk\sigma} g_{k+q,k}\,(u_{k+q}u_{k} - v_{k+q}v_k) \nonumber\\ 
& \times \gamma_{k+q,\sigma}^\dagger \gamma_{k\sigma} (b_q + b_{-q}^\dagger)	\nonumber\\
&+\sum_{qk\sigma} g_{k+q,k}\,(u_{k+q}v_{k} + v_{k+q}u_k) \nonumber\\ 
&\times  \gamma_{k+q,\sigma}^\dagger \overline \gamma^\dagger_{k\sigma} (b_q + b_{-q}^\dagger) + h.c.
\label{Hep}
\end{align}
It depends on the matrix element of the electron-phonon coupling, $g_{k+q,k}$, and 
the coherence factors 
\begin{eqnarray}
u_k^2=\frac{1}{2}\left(1+\frac{\xi_k}{E_k}\right),\;\;
v_k^2=\frac{1}{2}\left(1-\frac{\xi_k}{E_k}\right).
\end{eqnarray}
The latter relate the quasiparticle operators, $\gamma^\dagger_{k\sigma},\gamma_{k\sigma}$, to the electron operators in the dot, 
$d^\dagger_{k\sigma},d_{k\sigma}$, as follows
\begin{eqnarray}
\label{eq::Bogolioubov_transformation}
d_{k\sigma} &=& u_k\gamma_{k\sigma}+\sigma v_k\overline\gamma_{k\sigma}^\dagger,
\nonumber\\
d_{k\sigma}^\dagger &=& u_k\gamma_{k\sigma}^\dagger+\sigma v_k\overline\gamma_{k\sigma}.
\end{eqnarray}
Here we defined the "time-reversed" operators $\overline \gamma_{k\sigma} \equiv \gamma_{-k-\sigma}$ and assume
that $\sigma$ can take the values $\pm 1$ corresponding to spin up and down. 
The first sum in the Hamiltonian (\ref{Hep}) accounts for the inelastic scattering of quasiparticles on phonons and conserves the quasiparticle number, 
whereas the remaining terms describe Cooper pair breaking and quasiparticle recombination.

The last term in the Eq. (\ref{Ham}) is the sum of the tunnel Hamiltonians of the left and right junctions,
\begin{align}
 H_T &=\sum_{rkk'\sigma} t_{kk'}^r\,\hat T\, e^{-i \phi} c_{rk'\sigma}^\dagger (u_k\gamma_{k\sigma} + \sigma v_k \overline \gamma_{k\sigma}^\dagger) \\ \nonumber &+\sum_{rkk'\sigma} t_{kk'}^{r*}\,\hat T^\dagger \, e^{i \phi}  (u_k\gamma_{k\sigma}^\dagger + \sigma v_k \overline \gamma_{{k\sigma}}) c_{rk'\sigma}.
\end{align}
The operator $\hat T=\sum_n |n\rangle \langle n+1|$ accounts for changes of the number of electrons in the quantum dot, and
$\phi(t)=\int_{t_0}^t dt' \, eV_\phi(t')$ is the phase associated with the time-dependent gate voltage, with $t_0$ being an arbitrary initial time. 
To describe the experiment \cite{Maisi_2012} we will assume 
\begin{eqnarray}
V_\phi(t) = -\frac{eA_g}{C_g}\sin(2\pi ft),
\label{Vg}
\end{eqnarray}
which corresponds to harmonic pumping with the frequency $f$ and dimensionless amplitude $A_g$.

\subsection{Sequential tunneling approximation}
\label{section::sequential_tunneling}

We describe the dynamics of the system within the sequential tunneling approximation,
which is valid in the limit of weak tunneling 
\begin{eqnarray}
1/R_T^L, 1/R_T^R \ll e^2/(2\pi \hbar),
\label{condition1}
\end{eqnarray}
where $R_T^L,R_T^R$ are the resistances of the left and the right junctions.
We will further assume that the level spacing in the island is small compared to the temperature 
and the bias voltage and also that the frequency is smaller than the charging energy and the superconducting gap,  
$f \ll E_C,\Delta$. All these conditions were met in the single electron pumping experiment \cite{Maisi_2012}, 
which we are going to analyze in details below. 
Here we do not consider second order cotunneling contribution to the transport current, which may result in additional
tunneling events thus degrading the performance of the single electron turnstile. 
This contribution is obviously small in the limit (\ref{condition1}) and for small level spacing in the island \cite{Averin_2008,Averin_1992}, 
and it should be additionally suppressed by an exponential factor in an NSN structure due to the superconducting gap in the quasiparticle spectrum.

In second order perturbation theory in the tunnel Hamiltonian $H_T$, 
and within the Markov approximation, we obtain 
the master equation for the probabilities $p_n$ that the island of the SET transistor has excess charge $n$
 (see Appendix \ref{appendix::tunneling_rates} for details),  
\begin{eqnarray}
\label{eq::master_pn}
 \frac{d}{dt}p_n(t) &=& W_{n,n-1}(t)p_{n-1}(t)+W_{n,n+1}(t)p_{n+1}(t)
\nonumber\\ &&
-\, [W_{n-1,n}(t)+W_{n+1,n}(t)]p_{n}(t).
\end{eqnarray} 
The tunneling rates in this equation split into contributions from the left and the right junction, i.e.  
\begin{eqnarray}
W_{n+1,n}(t)= W_{n+1,n}^L(t)+W_{n+1,n}^R(t),
\end{eqnarray} 
where
\begin{align}
 W_{n+1,n}^r(t) &= \sum_{\sigma} \int d\xi\,  \bigg[ w_{n+1,n}^{r}(E,t)\frac{1-\mathcal{A}_nF_{\xi\sigma}}{2}\left(1+\frac{\xi}{E}\right)
\nonumber\\  &
+\, w_{n+1,n}^{r}(-E,t)\frac{\mathcal{A}_nF_{\xi\sigma}}{2}\left(1-\frac{\xi}{E}\right)\bigg],
\label{eq::rate_w}
\end{align}
and $E=\sqrt{\xi^2+\Delta^2}$.
The combination 
\begin{align}
w_{n+1,n}^{r}(E,t) = \frac{f^r\left(E_{n+1}-E_{n} +eV_\phi(t)-\mu_r + E\right)}{e^2R^r_T} 
\end{align} 
under the integral depends on the electron distribution
functions in the leads $f^r$ and the electrostatic energy $E_n=E_C(n-n_g^0)^2$
of the state with $n$ excess charges. 
The rates (\ref{eq::rate_w}) further depend on the quasiparticle distribution function in the superconducting island, via
\begin{eqnarray}
\mathcal{A}_n F_{\xi\sigma}=\frac{1}{{\mathcal N}_F{\mathcal V}} \sum_k \delta(\xi-\xi_{k}) \langle \gamma^\dagger_{k\sigma} \gamma_{k\sigma}\rangle_n.
\label{Fqp}
\end{eqnarray}
The expectation value $\langle\,\cdot\,\rangle_n$ in the right hand side of this equation
is taken at fixed number $n$ of electrons in the dot.
Since the level spacing in the island is assumed to be small, we may express this
average as the product of the "bulk" distribution function $F_{\xi\sigma}$, which is not sensitive
to the number of electrons in the dot, and the factor $\mathcal{A}_n$, which accounts for the parity effect
\cite{Averin_1992,cite:parity} 
(see Appendix \ref{app_parity} for details). This effect originates from the fact that
for even $n$ no quasiparticles exist in the ground state of the dot, while for odd $n$ at least one unpaired 
quasiparticle always remains excited. Having in mind the experiment \cite{Maisi_2012},
in the rest of this paper we will assume that 
$F_{\xi\sigma}\ll 1$ and that there is spin degeneracy in the problem, i.e. we assume $F_{\xi_k,\uparrow}=F_{\xi_k,\downarrow}$.
As we show in Appendix~\ref{app_parity} 
under these conditions one can express the parameter $\mathcal{A}_n$ as follows  
\begin{align}
\label{eq::factor_An}
\mathcal{A}_n = \left\{\begin{array}{cc}
                        \tanh(N_{qp}) & {\rm for}\;\;{\rm even}\;\; n \\
		        \coth(N_{qp}) & {\rm for}\;\;{\rm odd}\;\; n
                       \end{array}\right. ,
\end{align}
where
\begin{eqnarray}
N_{qp}=\sum_{k\sigma} F_{k\sigma}=\mathcal{N}_F\mathcal{V}\sum_\sigma\int d\xi F_{\xi\sigma}
\label{N_qp}
\end{eqnarray}
is the average number of excited quasiparticles in the superconducting dot provided
one would adopt a grand canonical approach to the problem and would allow the number of 
electrons in the dot to fluctuate. In Eq. (\ref{N_qp}) we have also defined
the density of states in the dot at the Fermi level $\mathcal{N}_F$ and the dot volume $\mathcal{V}$.  
Obviously in the limit of large number of quasiparticles, $N_{qp}\gg 1$, one finds ${\cal A}_n=1$ and
the parity effect vanishes. In the opposite limit $N_{qp}\ll 1$ we find $\mathcal{A}_n\to 0$ for even $n$ and 
$\mathcal{A}_n\to \infty$ for odd $n$.

In the same way we derive the remaining tunneling rates, which have the form
\begin{eqnarray}
W_{n-1,n}(t)= W_{n-1,n}^L(t)+W_{n-1,n}^R(t),
\end{eqnarray} 
where
\begin{align}
 W_{n-1,n}^r(t) &= \sum_{\sigma} \int d\xi\,  \bigg[ w_{n-1,n}^{r}(-E,t)\frac{1-\mathcal{A}_nF_{\xi\sigma}}{2}\left(1-\frac{\xi}{E}\right)
\nonumber\\  &
+\,w_{n-1,n}^{r}(E,t)\frac{\mathcal{A}_nF_{\xi\sigma}}{2}\left(1+\frac{\xi}{E}\right)\bigg],
\label{eq::rate_w1}
\end{align}
and
\begin{align}
w_{n-1,n}^{r}(E,t) 
=  \frac{1-f^r\left(E_{n}-E_{n-1}+eV_\phi(t) -\mu_r + E \right)}{e^2R^r_T}. 
\end{align} 

The master equation (\ref{eq::master_pn}) differs from the more familiar equation 
describing charge transport through an SET in two ways:
First, the tunneling rates in Eq. (\ref{eq::master_pn}) depend on time because of the sinusoidal modulation
of the gate voltage (\ref{Vg}). Second, the rates contain the distribution function of quasiparticles $F_{\xi\sigma}$,
which in  general differs from the equilibrium form.  
The time evolution of the latter is described by the following kinetic equation
\begin{align}
\label{eq::full_boltzmann}
& \frac{d}{dt}\bigg[\sum_n p_n \mathcal{A}_n F_{\xi\sigma} \bigg] = \frac{1}{\mathcal{N}_F\mathcal{V}} \sum_{rn}\sum_{s=\pm 1} p_n 
\nonumber\\ &\bigg[w_{n+s,n}^{r}(sE,t)\frac{1-\mathcal{A}_n F_{\xi\sigma}}{2}  
\left(1+s\frac{\xi}{E}\right) \nonumber \\ &-  w_{n+s,n}^{r}(-sE,t)\frac{\mathcal{A}_n F_{\xi\sigma}}{2}\left(1-s\frac{\xi}{E}\right) \bigg]
\nonumber \\ &
 +\, \pi \int d\xi'  b(E+E')^2 \left(1-\frac{\xi\xi'}{EE'}+\frac{\Delta^2}{EE'}\right)
\nonumber \\ &\times\,
\big[(1-F_{\xi\sigma})(1-F_{\xi'\bar\sigma})n^B_{E+E'} 
- F_{\xi\sigma}F_{\xi'\bar\sigma}(1+n^B_{E+E'}) \big]
\nonumber  \\ &
+\, \pi \int d\xi'  b(E'-E)^2{\rm sign}(E'-E)\left(1+\frac{\xi\xi'}{EE'}-\frac{\Delta^2}{EE'}\right)
\nonumber \\&\times  
\big[F_{\xi'\sigma}(1-F_{\xi\sigma})(1+n^B_{E'-E})-F_{\xi\sigma}(1-F_{\xi'\sigma})n^B_{E'-E} \big] \nonumber \\ &\times\sum_n p_n\mathcal{A}_n . 
\nonumber\\  
\end{align}
Here $\bar\sigma$ stands for the spin opposite to $\sigma$, and
$
n^B_\omega={1}/[e^{\omega/T}-1]
$
is the phonon equilibrium distribution function. 
Equation (\ref{eq::full_boltzmann}) has been derived in second order perturbation theory
in the Hamiltonians $H_T$ and $H_{e-ph}$, combined with the Markov approximation (see Appendix~\ref{appendix::kinetic_equation}).
The first three lines in its right hand describe the injection and leakage of the
non-equilibrium quasiparticles through the tunnel junctions. The fourth and fifth line contain the
terms responsible for the pairwise creation and annihilation of quasiparticles, and the last three
lines describe the scattering of quasiparticles on phonons. 
The parameter $b$ is expressed via the matrix element of electron-phonon coupling averaged
over the Fermi surface \eqref{def_b}. Finally we note that
 Eqs.~\eqref{eq::master_pn} and \eqref{eq::full_boltzmann} have a similar structure as those  
obtained  in Ref.~\cite{Averin_1990} for a normal conducting island. 

In general, the kinetic equation (\ref{eq::full_boltzmann}) should also contain the collision integral
induced by the short range Coulomb interaction between the electrons. In our model we omit it because
here we will mostly focus on the regime where the occupation probabilities of the
quasiparticle energy levels are small, $F_{\xi\sigma}\ll 1$. We will show below that in this limit the current through
our device depends only on the total number of quasiparticles $N_{qp}$, and not on the specific form of the
distribution function $F_{\xi\sigma}$. Since the electron-electron interaction does not cause 
recombination or creation of quasiparticle, it does not change $N_{qp}$ and, hence, may be ignored.
However, one should keep in mind that even at $F_{\xi\sigma}\ll 1$ the Coulomb interaction may change the
shape of the distribution function shown, i.e., in Fig. \ref{fig::charge_imbalance}(b).     

The kinetic equation (\ref{eq::full_boltzmann}) is the first main result of our paper.
It allows one to get access to the distribution function of quasiparticles and to study
how does it change under various bias conditions.  

We will now demonstrate that in the limit $F_{\xi\sigma}\ll 1$
one can replace the full kinetic equation (\ref{eq::full_boltzmann}) by a much simpler
equation for $N_{qp}$.  
The key point is that in the considered limit the  experimentally most relevant parameter -- the current flowing through the island --
can be expressed via $N_{qp}$. 
Indeed, in general the current through the junction $r$ is given 
by the sum (see Appendix~\ref{appendix::tunneling_current})
\begin{align}
\label{eq::current}
 I_r (t)= e \sum_n \big[W_{n+1,n}^r(t) - W_{n-1,n}^r(t)\big]\,p_n(t),
\end{align}
and at  $F_{\xi\sigma}\ll 1$ the tunneling rates $W_{n+1,n}^r(t),W_{n-1,n}^r(t)$ may be approximated as
\begin{align}
\label{eq::rates_Rothwarf_Taylor}
W_{n\pm 1n}^r&= \int d\xi w_{n\pm 1,n}^{r}(E) \nonumber \\
             &-[w_{n\pm 1, n}^{r}(\Delta)-w_{n\pm 1,n}^{r}(-\Delta)]\frac{\mathcal{A}_nN_{qp}}{2\mathcal{N}_F\mathcal{V}}.
\end{align} 
The dynamical equation for the quasiparticle number can be derived from the general equation (\ref{eq::full_boltzmann})
by taking the integral over $\xi$. In the section~\ref{section::quasiparticle_and_charge_imbalance} we will show that the charge imbalance in our system is small (for a quantitative discussion see  Fig.~\ref{fig::charge_imbalance}). Because of that one can put $F_{\xi\sigma}=F_{-\xi\sigma}$, which ultimately leads to  
the so called Rothwarf-Taylor equation \cite{Rothwarf_Taylor_1967} (see  Appendix~\ref{appendix::kinetic_equation} for details)
\begin{align}
 \frac{d}{dt}\left[N_{qp}\sum_n p_n \mathcal{A}_n \right]  
= \sum_n p_n\,\left[I_n^{qp} - \Gamma_n^{qp} N_{qp}\mathcal{A}_n - \kappa N_{qp}^2 \right].
\label{eq::Rothwarf_Taylor}
\end{align}  
Here 
\begin{eqnarray}
I^{qp}_n=\int d\xi \sum_r\big[ w_{n+1,n}^{r}(E,t) + w_{n-1,n}^{r}(-E,t)\big]
\end{eqnarray}
is the total quasiparticle injection rate,
\begin{eqnarray}
\Gamma^{qp}_{n}=\frac{1}{2\mathcal{N}_F\mathcal{V}}\sum_r \sum_{s=\pm 1} [w_{n+s,n}^{r}(\Delta) +  w_{n+s,n}^{r}(-\Delta)]
\end{eqnarray}
is the rate of tunneling of quasiparticles out of the dot, 
and $\kappa=4\Gamma_{e-ph}/\mathcal{N}_F\mathcal{V}\Delta$ characterizes the rate of quasiparticle recombination. It scales with
$\Gamma_{e-ph}=\pi b \Delta^3$, which is the characteristic time scale of electron-phonon scattering.  
The electron-phonon coupling constant $b$ can be related to experimentally more relevant parameter $\Sigma$, which appears in the 
heat current between electron to phonon subsystems in the normal state, $P_{e-ph}=\Sigma \mathcal{V}( T_e^5-T_{ph}^5)$. 
The corresponding relation reads  \cite{Giazotto_2006} 
\begin{align}
 b={\Sigma}/{48\pi\zeta(5)\mathcal{N}_F}.
\end{align}
For aluminum one has $\Sigma\approx 1.8\times10^9 \rm WK^{-5}m^{-3}$, $\Delta\approx 210\,\rm \mu eV$ and 
$\mathcal N_F\approx2.32\times 10^{28} \rm eV^{-1}m^{-3}$, which gives $\Gamma_{e-ph}\approx18\,\rm MHz$.
We note that at low temperatures the actual electron-phonon recombination rate is typically much smaller than $\Gamma_{e-ph}$, 
see Eq. (\ref{eq::recombination_rate}). 
For example, in the experiment \cite{Maisi_2012} it was found to be close to $10$ kHz.  

Eq. (\ref{eq::Rothwarf_Taylor}) is the second main result of our paper. As we have discussed, it is not sensitive
to the particular form of the distribution function $F_{\xi\sigma}$ and to the presence or the absence
of short range electron-electron interaction. Besides that, it is much easier to
solve than the full kinetic equation (\ref{eq::full_boltzmann}).
We would also like to note that at low temperatures Eq. (\ref{eq::Rothwarf_Taylor})
leads to the same results as the formalism used in Ref. \cite{Maisi_2012}. We discuss this point in more detail 
in Appendix \ref{QP_rates}.

\section{Results and Discussion}
\label{sec::results}

\begin{table}[Iht]
 \begin{ruledtabular}
  \begin{tabular}{ccccc}
$E_C$ &
$eV$ & 
$T$ &
$(e^2R_T^L\mathcal{N}_F \mathcal{V})^{-1}$ &
$(e^2R_T^R\mathcal{N}_F \mathcal{V})^{-1}$  \\
$8\Delta/7$ & 
$4\Delta/3$ &
$\Delta/40$ &
$1.8 \times 10^{-2}\Gamma_{e-ph}$ &
$2.5\times 10^{-2}\Gamma_{e-ph}$
  \end{tabular}
 \end{ruledtabular}
 \caption{Parameters used in the simulations unless other values are specified. 
The frequency $\Gamma_{e-ph}=\pi b \Delta^3$ gives a characteristic scale for the rate of the electron-phonon relaxation. 
The chosen  parameters produce the best fit to the experimental data of Ref.~\cite{Maisi_2012}.}
 \label{table::parameters}
\end{table}

\begin{figure}[!ht]
	\begin{center}
	 \includegraphics[width=0.8\columnwidth]{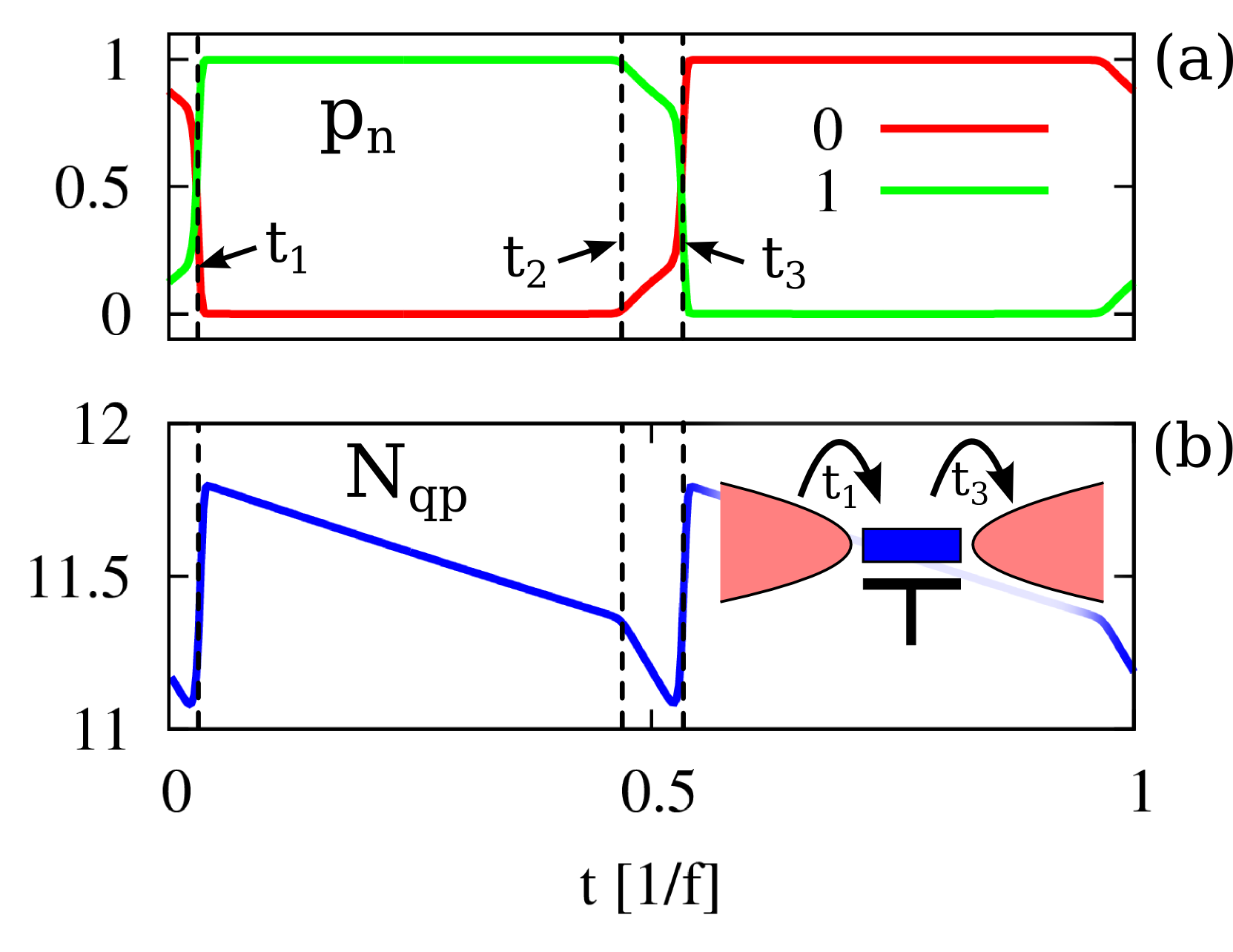} 
\caption{(a) Time evolution of the occupation probabilities of the charging states
with $0$ ($p_0$) and $1$ ($p_1$) extra electrons in the island. The gate
voltage is modulated according to  Eq. (\ref{Vg}) with 
frequency  $f=5.56\times 10^{-2}\Gamma_{e-ph}$. Other system parameters are listed in 
 Table~\ref{table::parameters}.
(b) Time evolution of the quasiparticle number $N_{qp}$. The inset illustrates the different tunneling events.}
\label{fig::system}
\end{center}
\end{figure}

The system of two coupled equations \eqref{eq::master_pn} and \eqref{eq::full_boltzmann} can be
readily solved numerically, yielding the full information about the distribution function in the
quantum dot and all other parameters. As we have already mentioned we assume spin degeneracy, so that $F_{\xi\sigma}=F_{\xi\bar\sigma}$.

Let us first consider the regime of large number of excited quasiparticles, $N_{qp}\gg 1$ and $\mathcal{A}_n \approx 1$. We find 
the latter to be approximately fulfilled for $N_{qp}\gtrsim 2$ [see Fig.~\ref{fig::appendix_Nqp}]. In this limit
the parity effect is negligible.

In Fig.~\ref{fig::system} (a) and (b) we 
show the time-dependence of the occupation probabilities of the charging states $n=0$ and $n=1$ together with the 
quasiparticle number in the dot, $N_{qp}$, for the set of parameters listed in Table~\ref{table::parameters}.
The sinusoidal modulation of the gate voltage allows for different tunneling processes in certain time windows, defining the times $t_{i}, i=1,2,3$, 
all depending on the modulation amplitude of the gate voltage.  At $t_1$ an electron can tunnel (and does so nearly immediately once it is allowed) from the left lead to the superconducting quantum dot, changing the charge state from $n=0$ to $n=1$.
Because of the Coulomb blockade no further single-electron tunneling occurs.
The tunneling process also increases the quasiparticle number, which in the following relaxes back  due to recombination with  rate (which will be further discussed in the next section)
\begin{eqnarray}
\frac{1}{\tau_{rec}}=\kappa N_{qp}.
\end{eqnarray} 
Beyond the time $t_2$ quasiparticles may also escape to the leads via tunneling
within the time interval $(t_2,t_3)$.
The corresponding escape-rate is given by 
$$
\Gamma^r_{tun}=\frac{1}{2e^2R^r_T\mathcal{N}_F\mathcal{V}}.
$$ 
Next, at a time $t_3$ an electron leaves the dot through the right junction, and the cycle of processes repeats.  
One can see that between the times $t_1$ and $t_3$ one electron charge has been pumped through the system from the left to the right.
It is also interesting to note that the decay of $N_{qp}$ during the time interval $t_2<t<t_3$ 
and its rapid rise at time $t_3$ sum up to 1, which is the total change of the
electron number in the dot in the same period of time.
Thus one can say that at $t_2<t<t_3$ the number of electron-like quasiparticles decreases, while at $t=t_3$
the number of hole-like quasipartilces rises.

\subsection{Quasi-particle number and charge imbalance}
\label{section::quasiparticle_and_charge_imbalance}

\begin{figure}[!ht]
\begin{center}
\includegraphics[width=1\columnwidth]{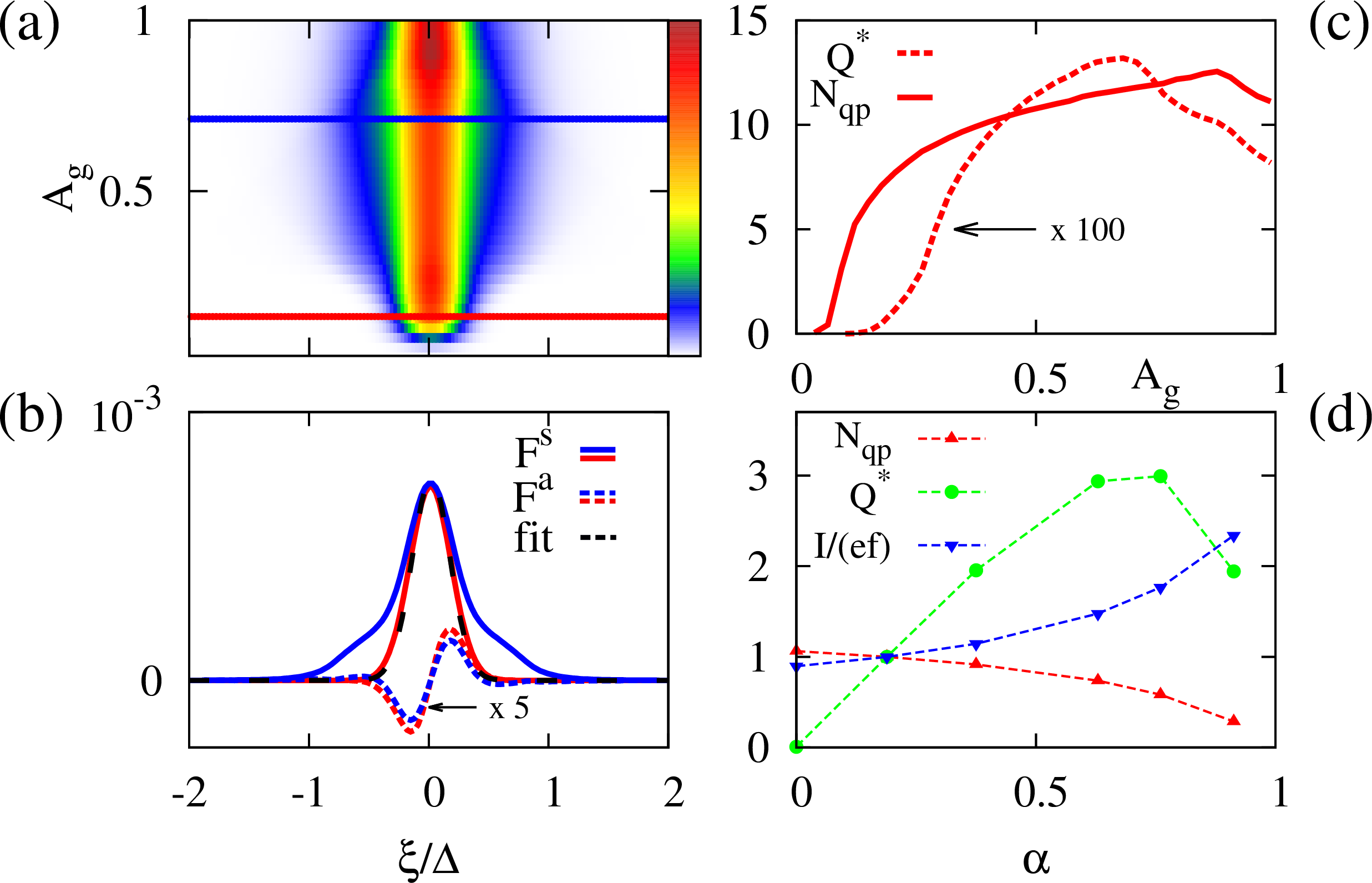} 
\end{center}	
\caption{ Time averaged distribution function $\langle F_{\xi}\rangle$ and quasiparticle imbalance  charge $\langle Q^* \rangle$ 
at the modulation frequency $f=5.56\times10^{-2}\,\Gamma_{e-ph}$.  
(a) Color plot of $\langle F_{\xi}\rangle$ versus energy $\xi$ and gate modulation amplitude $A_g$. 
(b) Symmetric ($\langle F_\xi^s \rangle$) and asymmetric ($\langle F_\xi^a \rangle$) components of the distribution function along the red and blue cuts 
in the panel (a). The dashed black line shows the approximate function $F^{\rm loc}_E$ in (\ref{Floc}). 
(c) Dependence of the time averaged quasiparticle number $\langle N_{qp} \rangle$ and the quasiparticle charge density $\langle Q^* \rangle$
on the pumping amplitude $A_g$. 
(d) Dependence of $\langle I/(ef)\rangle$, $\langle N_{qp} \rangle$ and $\langle Q^* \rangle$ on the junction asymmetry parameter $\alpha=(R_T^R-R_T^L)/(R_T^R+R_T^L)$. The quantities are evaluated for $A_g=0.5$ and normalized by the results obtained at $\alpha=0.18$. }
\label{fig::charge_imbalance}
\end{figure}

Figs.~\ref{fig::charge_imbalance} (a) and \ref{fig::charge_imbalance} (b) show the time averaged quasiparticle distribution function
\begin{eqnarray}
\langle F_{\xi}\rangle = f\int_{-1/2f}^{1/2f} dt F_{\xi}(t).
\end{eqnarray} 
We observe that the function $\langle F_{\xi}\rangle$ deviates from the equilibrium form.
First of all, it becomes slightly asymmetric in $\xi$, i.e. $\langle F_{\xi}\rangle\not=\langle F_{-\xi}\rangle$.
The degree of this asymmetry is controlled by the asymmetry in the junction resistances $R_T^r$.   
Second, its value at $\xi=0$ is increased compared to what one finds in thermal equilibrium. 

Traditionally the distribution function is decomposed into a symmetric and asymmetric part $F_\xi^{s,a}=(F_{\xi} \pm F_{-\xi})/2$ \cite{quasiclassical,Kenneth_Gray_1981}.
They determine, respectively, the quasiparticle number $N_{qp}$ (\ref{N_qp}) and the quasiparticle charge density $Q^*$ \cite{Clarke_Tinkham_1972},
which is given by the integral
\begin{align}
\label{eq::qp_number}
 Q^*=\mathcal{N}_F \mathcal V \int d\xi \,\frac{\xi}{E} \,F^a_{\xi}.
\end{align}
Comparing $Q^*$ to $N_{qp}$ one can draw a conclusion about the magnitude of charge imbalance induced in the quantum dot. 
Both quantities are presented in Fig.~\ref{fig::charge_imbalance} (c) as functions of the gate modulation amplitude $A_g$. 
Note that the charge imbalance $Q^*$ is much smaller compared to the quasiparticle number $N_{qp}$, which assures the use of Eq.~\eqref{eq::Rothwarf_Taylor} in order to describe the quasiparticle kinetics on the island.
Fig.~\ref{fig::charge_imbalance} (d) demonstrates that both $Q^*$ and the current increase with the asymmetry in the resistances of the two junctions, 
whereas the quasiparticle number decreases with the asymmetry.

One can get more insight into the results of the numerical simulations if one analyses the interplay between
two different channels of quasiparticle relaxation, namely the inelastic phonon scattering and  recombination. 
The corresponding relaxation rates, $1/\tau_{sc}$ and $1/\tau_{rec}$, can be derived from the kinetic equation
\eqref{eq::full_boltzmann}. Neglecting for the moment the charge imbalance we approximately find the rates in the form
\begin{align}
 \frac{1}{\tau_{sc}(E)}&= \Gamma_{e-ph}\frac{\sqrt{2}(E-\Delta)^{7/2}}{E\Delta^{5/2}}, \label{eq::scattering_rate} \\
 \frac{1}{\tau_{rec}(E)}&= \Gamma_{e-ph}\frac{(E+\Delta)^3}{2E\Delta^2} \frac{N_{qp}}{\mathcal{N}_F \mathcal V\Delta}, \label{eq::recombination_rate}
\end{align}
which are consistent with the temperature dependent inverse lifetimes derived in Ref.~\cite{Kaplan_1976}.
At energy $E_0\approx\Delta(1+2({N_{qp}}/{2{\mathcal N}_F \mathcal V\Delta})^{2/7})$ these rates are equal,
at $E>E_0$ the inelastic scattering dominates, while at $E<E_0$ the recombination becomes more important. 
At $N_{qp}\ll 1$ the energy $E_0$ is close to $\Delta$, which leads to the following scenario: high
energy quasiparticles are quickly equilibrated by inelastic phonon scattering and the resulting quasi-equilibrium
distribution with the phonon temperature subsequently slowly decays due to recombination until this decay is balanced by the influx
of new quasiparticles from the leads. Thus, within this simple model the distribution function should have 
a local equilibrium form 
\begin{eqnarray}
F_E^{\rm loc}=[\exp((E-\mu)/T)+1]^{-1}.
\label{Floc}
\end{eqnarray}
where an increase in the quasiparticle number is expressed as a shifted chemical potential $\mu$.
To avoid confusion at this point, we note that $\mu$ is not related to charge imbalance (we actually ignored it), 
it merely indicates an increased number of quasiparticles.     
A fit of the distribution function to Eq. \eqref{Floc} along the red cut 
in Fig.~\ref{fig::charge_imbalance} (a) approximately yields $\mu=0.82\,\Delta$. 
The fit is plotted by the black dashed line in Fig.~\ref{fig::charge_imbalance} (b)
and turns out to be very good. 
A similar scenario of relaxation of non-equilibrium quasiparticles 
had been discussed a long time ago by Owen and Scalapino~\cite{Owen_Scalapino_1972}.

As we have discussed above, the precise form of the distribution function $\langle F_\xi\rangle$
is not important as long as $\langle F_\xi\rangle\ll 1$ and one is only interested in the current
flowing through the device. It becomes important, however, if one is interested in more
subtle effects like relaxation, excitation or decoherence of the quantum states of qubits \cite{Wenner}.
As we have demonstrated in this section, our model may be useful in describing such phenomena.

\subsection{Frequency dependence}
\label{section::frequency_dependence}

\begin{figure}[!ht]
	\begin{center}
	 \includegraphics[width=\columnwidth]{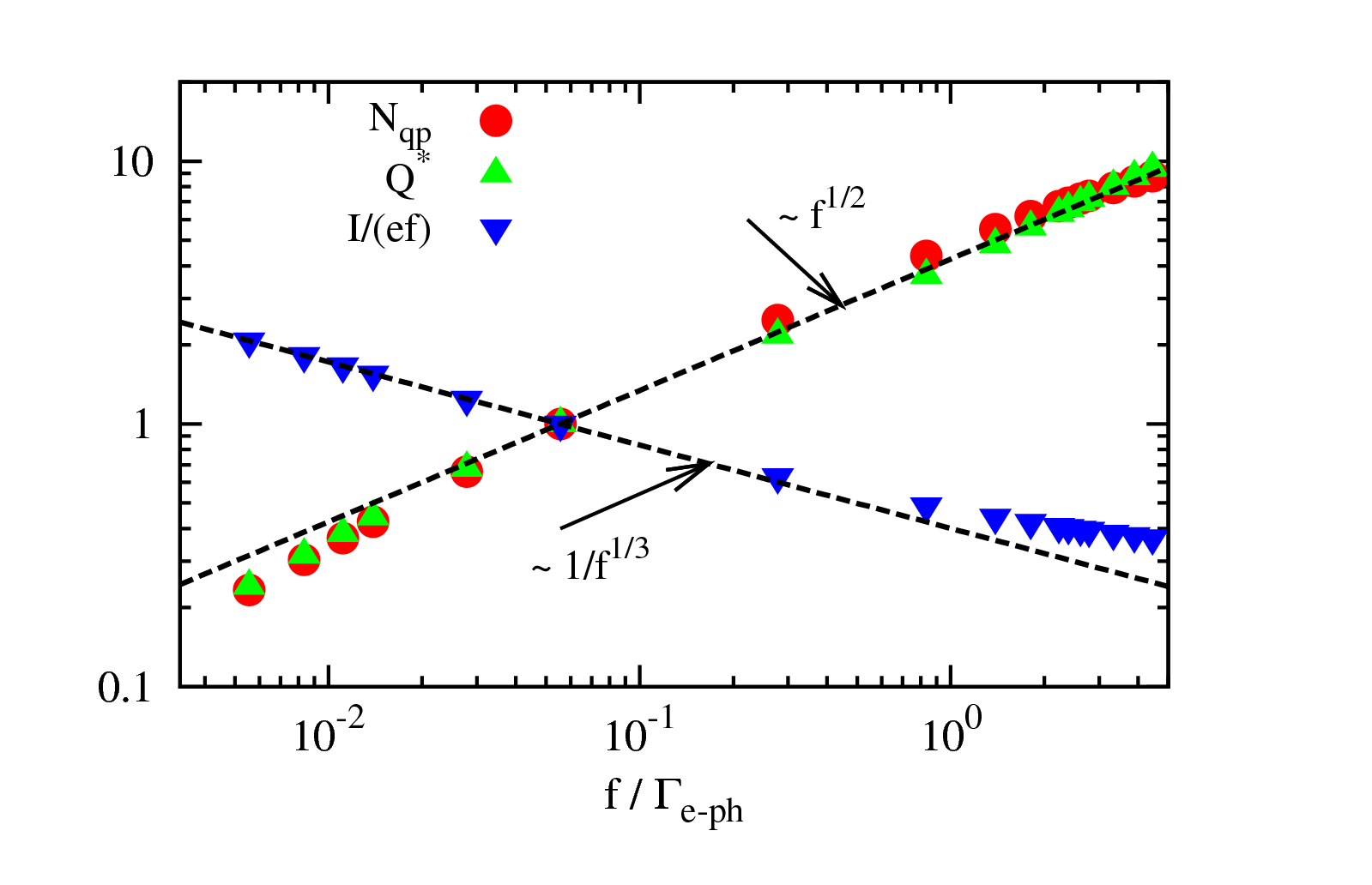} 
	\end{center}
\caption{Frequency dependence of the quasiparticle number $\langle N_{qp} \rangle $, the quasiparticle charge density $\langle Q^* \rangle $ and the normalized pumping current $\langle I/(ef)\rangle $. 
The quantities are evaluated for a gate amplitude $A_g=0.5$ and normalized by the results obtained at $f=5.56\times10^{-2}\, \Gamma_{e-ph}$.
	 }
	 \label{fig::frequency_scaling}
\end{figure}

\begin{figure}[!ht]
	\begin{center}
$
\begin{array}{c}
	 \includegraphics[width=\columnwidth]{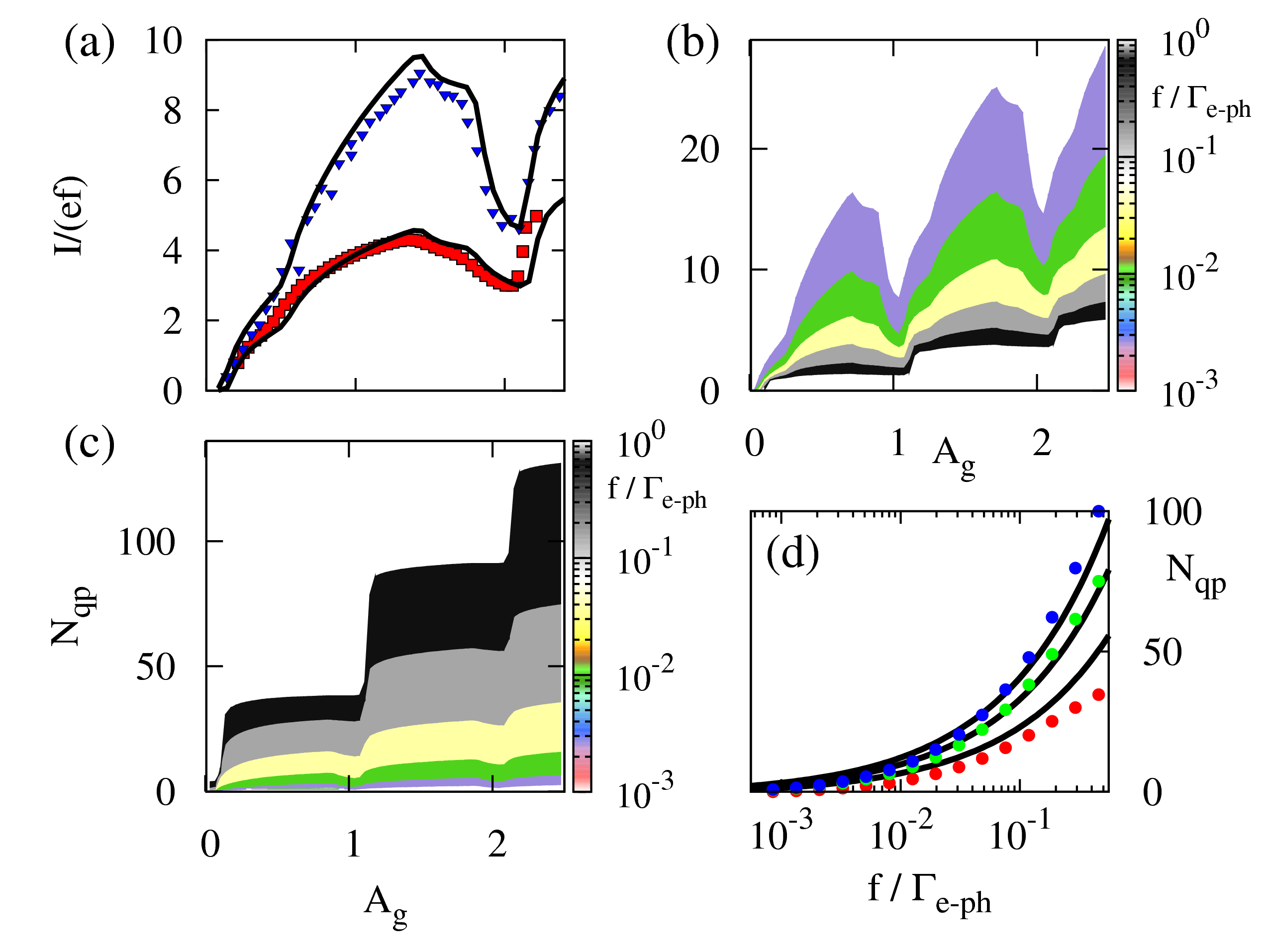} \\
\end{array}
$
\caption{Normalized current $\langle I  /ef\rangle$ and the quasiaprticle number $\langle N_{qp}\rangle $ at different modulation amplitudes and frequencies. 
(a) Normalized current, $\langle I  /ef\rangle$, versus the modulation amplitude $A_g$ for two different frequencies 
$f=5.56 \times 10^{-2}\Gamma_{e-ph}$ (red symbols) and $f=5.56 \times 10^{-3}\Gamma_{e-ph}$ (blue symbols);  
black line --- theory,  symbols --- experimental data of Ref.~\cite{Maisi_2012}. 
(b) Simulated current for a wide range of frequencies. 
(c) Quasiparticle number $\langle N_{qp} \rangle$ at various pumping frequencies and modulation amplitudes. 
(d) Frequency dependence of the quasiparticle number $\langle N_{qp}\rangle $ evaluated at the modulation amplitudes corresponding to the middle 
of plateaus in panel (c). The red, green and blue dots correspond, respectively,  to the first, second and third plateaus. 
The black lines indicate the approximation $\langle N_{qp} \rangle=\sqrt{2Nf/\kappa}$, where $N$ is the number of the plateau.}
	 \label{fig::current_exp_1}
	\end{center}
\end{figure}

An important question in the context of metrology and quantum information is the dependence of the quasiparticle poisoning 
of superconducting devices on the repetition rate with which an operation is performed. It is known that 
the operation frequency of the hybrid turnstile, which we are considering, should be chosen sufficiently low, $2\pi f < \Delta/e^2R_T$ \cite{Pekola_2008}, 
in order to leave electrons enough time to tunnel through the device. In our simulations we will stay below this high-frequency limit 
paying more attention to limitations of the device operation at low frequencies. 

In Fig.~\ref{fig::frequency_scaling} we investigate the frequency dependence of $\langle N_{qp} \rangle $, $\langle Q^*\rangle $ and $\langle I/(ef) \rangle$ for our setup. 
We find that both quasiparticle number and the charge density are determined by the recombination rate at high frequency. 
Indeed, at large $f$ and, hence, large $N_{qp}$ the term $\kappa N_{qp}^2$ on the right hand side of the Eq.~\eqref{eq::Rothwarf_Taylor} 
dominates over the term $\Gamma_n^{qp}N_{qp}\mathcal{A}_n$.  
The injection term $I_n^{qp}$ scales linearly with the frequency $f$ in this regime. 
Thus at high $f$ and after time averaging, Eq. \eqref{eq::Rothwarf_Taylor} leads to the result $\langle N_{qp} \rangle \propto {f}^{1/2}$. 
We find that for frequencies $f \gtrsim 0.1\,\Gamma_{e-ph}$ this dependence agrees with the numerical results fairly well. 
The normalized current $\langle I/(ef)\rangle$ tends to a constant in this limit, which makes it interesting for metrological applications. 
This limiting behavior of the current can be easily understood if one analyses the dependence of the rates \eqref{eq::rates_Rothwarf_Taylor}, 
which enter the current \eqref{eq::current}, on frequency. 
The first contribution to the rates  \eqref{eq::rates_Rothwarf_Taylor} scales as $\sim f$, while the quasiparticle contribution is proportional to 
$\langle N_{qp} \rangle$ and therefore it is suppressed at large frequencies. 
In the opposite limit of low frequency,  $f \lesssim 0.1\,\Gamma_{e-ph}$, we find that the 
numerical results are well fitted by the dependence $\langle I/(ef) \rangle \propto 1/f^{1/3}$.
 
Next we fit our model to the experimental data for the pumping current \cite{Maisi_2012},  
with the results shown in Fig.~\ref{fig::current_exp_1}. The theoretical curves have been generated by solving
the Rothwarf-Taylor equation~\eqref{eq::Rothwarf_Taylor} for $N_{qp}$ numerically and substituting the result in the expression for the current~\eqref{eq::current}. 
We find good agreement between theory and experiment, see Fig.~\ref{fig::current_exp_1} (a). 
Figs.~\ref{fig::current_exp_1} (b) and (c) show theory predictions for the current and the quasiparticle number, respectively, at various 
frequencies and gate modulation amplitudes. 
We find that at higher frequencies the normalized current $\langle I/(ef) \rangle$ approaches the ideal staircase-like behavior in agreement with our previous discussion. 
We also find that the quasiparticle number grows both with the frequency and with the gate modulation amplitude. 
This behavior can be readily understood if one returns to the  
time traces in Figs.~\ref{fig::system} (a) and \ref{fig::system} (b). 
Obviously, electron tunneling in or out of the dot is always accompanied by quasiparticle injection. 
As the gate modulation amplitude $A_g$ grows, a third charging state of the dot becomes available for the transport at some point, and the number of excited quasiparticles per cycle doubles. Assuming that exactly one quasiparticle is excited in every tunneling event we arrive at a simple estimate 
$\langle N_{qp} \rangle=\sqrt{2Nf/\kappa}$, where $N$ stands for the number of the plateau in Fig.~\ref{fig::current_exp_1} (c). 
In Fig.~\ref{fig::current_exp_1} (d) $\langle N_{qp} \rangle$ is plotted as a function of frequency $f$ 
and at $A_g=N-0.5$ (with $N=1,\,2,\,3$ corresponding to the red, green and blue dots).  
We observe that our simple estimate of the quasiaprticle number actually works
reasonably well. 

We would like to emphasize that we have only two independent fit parameters in our model, namely the
combinations $e^2 R_T^r {\cal N}_F{\cal V}$, $r=L,R$, which are listed in Table I. This observation provides
strong evidence of the validity of Eq. (\ref{eq::Rothwarf_Taylor}).

\subsection{Parity Effect} 
\label{section::parity_effect}

\begin{figure}[!ht]
	\begin{center}
	 \includegraphics[width=\columnwidth]{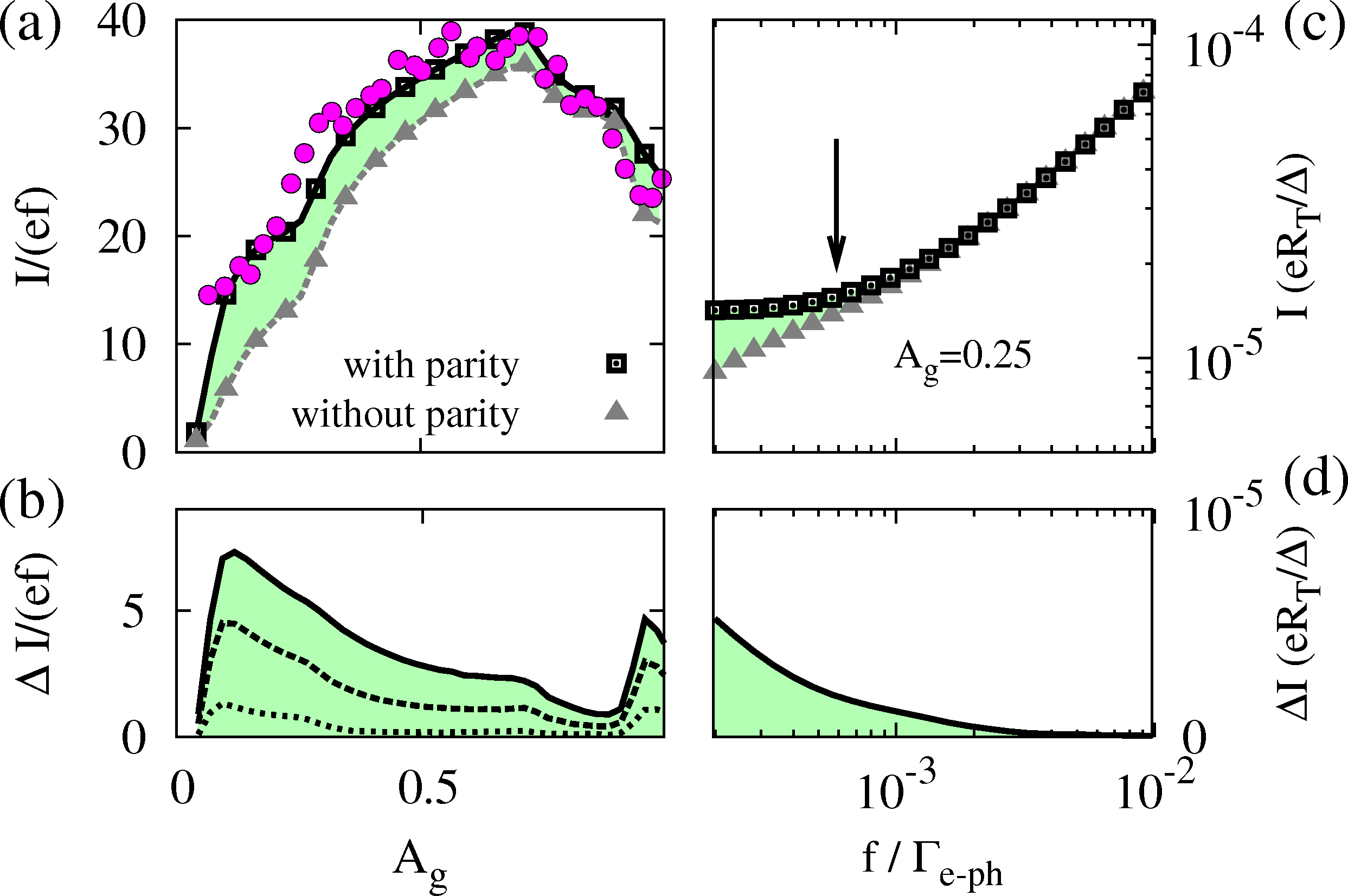} 
	\end{center}
	\caption{ The influence of the parity effect on the pumping current. 
(a) Theoretical current with parity effect included (black rectangles) and ignored (gray triangles) compared with the
experimental data of Ref.~\cite{Maisi_2012} (magenta circles) for $f=2\times 10^{-4} \Gamma_{e-ph}$. 
(b) Difference of the current with parity effect included and the one without it for the frequencies $f=2,\;2.7,$ and 
$5.4 \times 10^{-4} \Gamma_{e-ph}$ (solid,dashed and dotted lines respectively).
(c) Frequency dependence of the current with/without parity effect included for $A_g=0.25$ (black rectangles/gray triangles). The arrow indicates the recombination rate for $N_{qp}=2$.
(d) Difference of the curves in panel (c).}
	 \label{fig::parity}
\end{figure}

Let us now turn to the regime where the parity effect becomes important. 
In Fig.~\ref{fig::setup_stability_diag}(b) we calculated the stability diagram in equilibrium for $T=\Delta/40$. There the presence of an extra quasiparticle excitation in the odd charging state leads to a finite current plateau at $|eV| < 2\Delta$, which is 2-e-periodic in the gate charge. 

Due to the periodical excitation of quasipartciles during the turnstile operation the parity effect matters if the frequency becomes lower than the recombination rate  \eqref{eq::recombination_rate}, $f\lesssim 1/\tau_{\rm rec}(\Delta)$. 
In this case one finds that $N_{qp}\lesssim 2$, which is precisely the regime where the parity effect has to be taken into account (see Fig.~\ref{fig::appendix_Nqp}).
 
To see its influence on the average current we first solve Eqs.~\eqref{eq::master_pn} and \eqref{eq::Rothwarf_Taylor} in combination with
 Eq.~(\ref{eq::factor_An}). We compare the result of this full analysis with the simplified approach, in which we deliberately set 
$\mathcal{A}_n=1$ everywhere thus suppressing the parity effect.  
In Fig.~\ref{fig::parity} (a) we compare results of both approaches for the pumping current as a function of the modulation amplitude $A_g$ with the experimental data of Ref.~\cite{Maisi_2012}. 
We find that incorporating the parity effect into the model indeed allows us to better fit the experimental data, especially for small values
of the gate modulation amplitude. 
The role of the parity effect may be characterized by the difference between the exact current, $\langle I\rangle$,
and its value $\langle I^*\rangle$ derived under the assumption that $\mathcal{A}_n=1$.
In Fig.~\ref{fig::parity}(b) this difference, $\langle \Delta I \rangle=\langle I\rangle-\langle I^*\rangle$ is plotted as a function of the modulation amplitude for three different modulation frequencies showing that features of the parity effect first develop at small gate amplitude.  
In Fig.~\ref{fig::parity}(c) and \ref{fig::parity}(d), the frequency dependence of the currents $\langle I \rangle $ as well as $\langle I^*\rangle$ is shown for $A_g=0.25$. We find that the exact current $\langle I \rangle$ approaches a constant value for frequencies $f \lesssim 1/\tau_{\rm rec}(\Delta)$ whereas $\langle I^*\rangle$ decreases further with frequency. This agrees with the observation that in this regime the dominant current contribution arises from the current plateau that we see in Fig.~\ref{fig::setup_stability_diag}(b). Finally in panel Fig.~\ref{fig::parity}(d) the difference $\langle \Delta I \rangle$ is plotted versus frequency showing the emergence of the parity effect with decreasing frequency.

\section{Summary}
\label{sec::summary}

We have investigated the properties of a small superconducting island in a NSN configuration
driven by both a dc bias voltage and an ac pumping gate voltage. Apart from the number of excess single-electron charges on the dot we have to pay attention to the non-equilibrium distribution of quasiparticles.
Starting from the microscopic theory and using standard approximations we derived the master equation for the occupation probabilities of different charge states of the dot (\ref{eq::master_pn}).
The tunneling rates, which appear in this equation, are influenced by the 
non-equilibrium quasiparticle distribution function.
In addition we derived the kinetic equation
describing the time evolution of the quasiparticle distribution function (\ref{eq::full_boltzmann}).
The combination, i.e., Eqs.~(\ref{eq::master_pn}) and (\ref{eq::full_boltzmann}) fully describe the
dynamics of our system. We solved these equations numerically and demonstrate that
our model allows fitting with high precision the results of the experiment \cite{Maisi_2012}.
We have also derived a simplified kinetic equation (\ref{eq::Rothwarf_Taylor}) (Rothwarf-Taylor equation), 
which involves the total number of excited quasiparticles instead of their full distribution function.
This equation is valid in the regime where the occupation probabilities of the quasiparticle levels
are small. Our theory is valid even in the regimes where only one quasiparticle is excited in the superconducting
island. In particular, it fully takes into account the parity effect, which becomes important in small
superconducting particles at low temperatures.\cite{Averin_1992,cite:parity}

\vspace{3 mm}
{\bf Acknowledgements}

We thank D.V. Averin, F.W.J. Hekking, T. Heikkil\"a and J. Cole
for useful discussions. The work has been supported partially by LTQ (project
no. 250280) CoE grant and the National Doctoral Programme
in Nanoscience (NGS-NANO).

\appendix

\section{Density matrix at fixed number of electrons in the island}

\label{app_parity}

As usual, we separate the Hamiltonian into an unperturbed part $H_0=H_D+\sum_r H_r + H_p$ and a perturbation $H_I=H_{e-ph}+H_{T}$. 
After standard manipulations and making the Markov approximation we arrive at 
the Liouville equation for the density matrix of the system $\hat\rho(t)$,
\begin{align}
 \label{eq_appendix::liouville}
 \frac{d}{dt}\hat\rho(t) &= {(-i)^2}\int_{-\infty}^t dt' \, [H_I(t),[H_I(t'),\hat \rho(t)]].
\end{align}
Next we assume that the density matrix can be factorized into the product of the density matrices
of the leads $\hat\rho_r$, of the phonons, $\hat\rho_p$, and of the island. We also assume that the leads and
the phonons remain in equilibrium, so that  
\begin{align}
 \hat \rho_r=\frac{e^{-\beta H_r}}{{\rm Tr}(e^{-\beta H_r})},\quad \hat \rho_p=\frac{e^{-\beta H_p}}{{\rm Tr}(e^{-\beta H_p})},
\end{align}
where $\beta=1/T$ is the inverse temperature. 

Assuming a grand canonical ensemble, i.e. allowing fluctuations of the number of electrons in the dot, 
we may express the density matrix of the quasiparticles in the form 
\begin{align}
  \label{eq_appendix::qp_density_matrix}
 \hat \rho_{qp} = \prod_{k\sigma} \left[(1-F_{k\sigma})(1-\hat n_{k\sigma}) + F_{k\sigma} \hat n_{k\sigma} \right],
\end{align}
where $\hat n_{k\sigma}=\gamma^\dagger_{k\sigma} \gamma_{k\sigma}$ and $F_{k\sigma}$ is the occupation probability of the level $k\sigma$. 
In equilibrium one finds $F_{k\sigma}=1/(\exp(\beta E_k)+1)$. 
Out of equilibrium $F_{k\sigma}$ has to be obtained from a kinetic equation. 

Next, we include the parity effect into the model. Quite generally, one would have to switch to the
canonical ensemble and strictly fix the number of electrons in the dot, but this route
turns out to be technically very difficult. Fortunately, in order to describe the 
properties of big superconducting quantum dots with small level spacing and large number of electrons
it is sufficient to fix only the parity of the electron number. 
In order to do so we first introduce the projection operators $\hat P^\pm$
on the subspaces with even (denoted by the superscript +) and odd (denoted by the superscript -) numbers of
electrons trapped in the quantum dot. These operators read
\begin{align}
 \hat P^\pm &=\frac{1}{\sqrt{2}}\left[1\pm(-1)^{\hat n}\right]=\frac{1}{\sqrt{2}}\bigg[1 \pm  \prod_{k\sigma} (1-2\hat n_{k\sigma}) \bigg].
\end{align}
Here we have used the fact that the parities of the number of 
electrons in the quantum dot and of the number of excited quasiparticles are the same and made the replacement
\begin{eqnarray}
\hat n = \sum_{k\sigma} d^\dagger_{k\sigma}d_{k\sigma} \to \sum_{k\sigma} \gamma^\dagger_{k\sigma}\gamma_{k\sigma}. 
\end{eqnarray} 
Besides that, we also used the identity $(-1)^{\gamma^\dagger_{k\sigma}\gamma_{k\sigma}} = 1-2\gamma^\dagger_{k\sigma}\gamma_{k\sigma}$.
With the aid of the operators $\hat P^\pm$ we can write the density matrices of the
quasiparticles in the even and odd states in the form 
\begin{align}
 \hat \rho_{qp}^{n=even/odd} &= \frac{\hat P^\pm \hat \rho_{qp}}{{\rm Tr} (\hat P^\pm \hat \rho_{qp})}.
\end{align}
We further assume that the function $F_{k\sigma}$ remains the same in the even and odd states.
This assumption is valid in big quantum dots where one additional electron does not significantly
change the occupation probabilities of the energy levels.

\begin{figure}[!ht]
 \begin{center}
  \includegraphics[width=0.8\columnwidth]{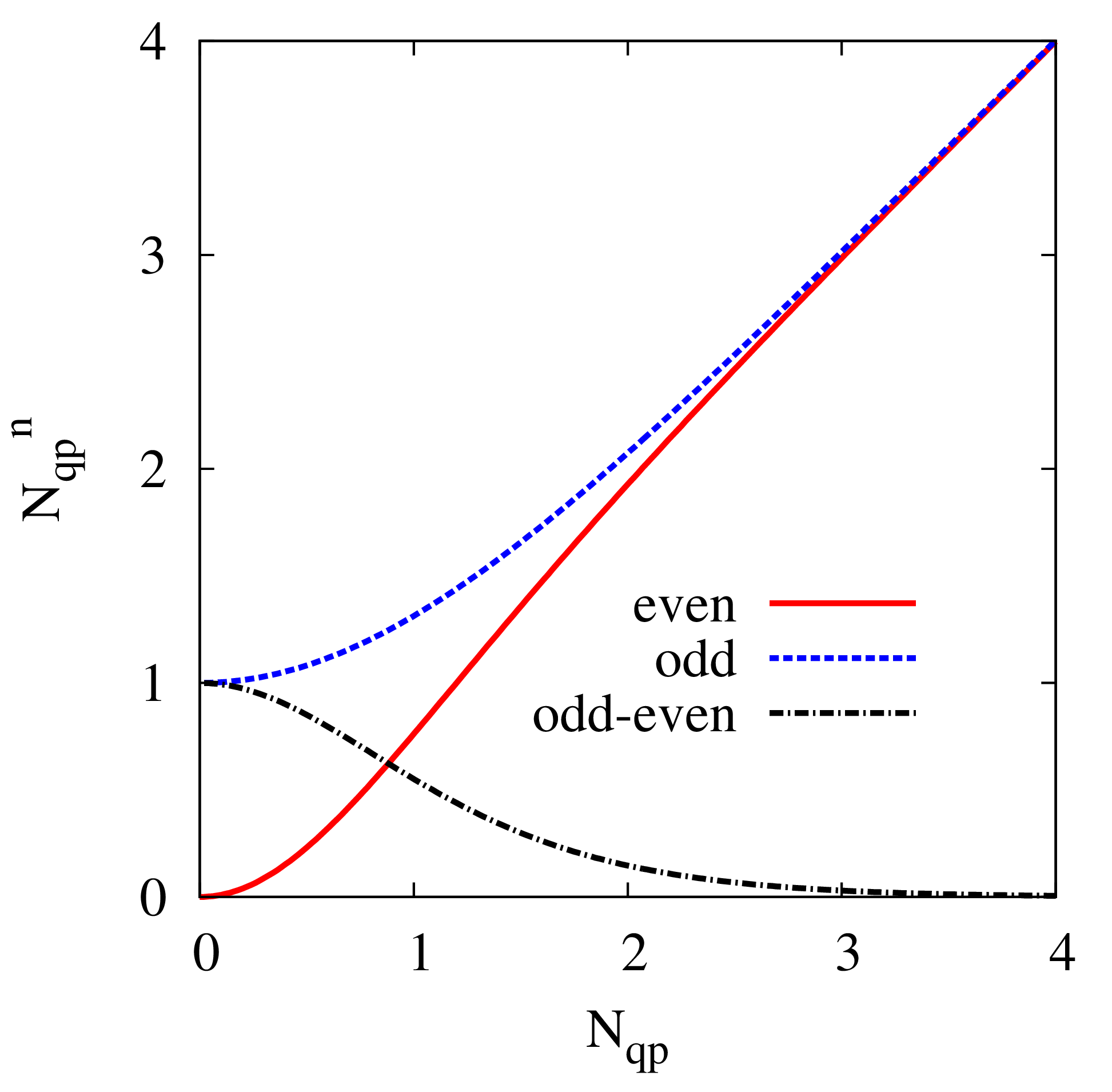}
 \end{center}
 \caption{ Parity affected quasiparticle number $N_{qp}^n=\sum_{k\sigma} \langle \hat n_{k\sigma} \rangle_{n}$ 
in the even and odd charging states versus the grand canonical quasiparticle number $N_{qp}$.}
  \label{fig::appendix_Nqp}
\end{figure}

The quasiparticle occupation probabilities $\langle \hat n_{k\sigma} \rangle_{n}={\rm Tr} (\hat n_{k\sigma} \hat \rho_{qp}^{n})$ can be readily calculated. 
Employing the commutation rules
{\small
\begin{align}
\label{eq_appendix::projection_operator}
 \gamma_{k\sigma} \hat P^\pm &=  \frac{1}{\sqrt{2}}\bigg[1 \mp  \prod_{p\alpha\neq k\sigma} (1-2\hat n_{p\alpha}) \bigg]\gamma_{k\sigma}, \\
 \gamma_{k\sigma}^\dagger \hat P^\pm &= \frac{1}{\sqrt{2}}\bigg[1 \pm  \prod_{p\alpha\neq k\sigma} (1-2\hat n_{p\alpha}) \bigg]\gamma_{k\sigma}^\dagger , \nonumber\\
 \hat n_{k\sigma} \hat P^\pm &=   \frac{1}{\sqrt{2}}\bigg[1 \mp  \prod_{p\alpha\neq k\sigma} (1-2\hat n_{p\alpha}) \bigg]\hat n_{k\sigma},\nonumber\\
(1- \hat n_{k\sigma}) \hat P^\pm &=   \frac{1}{\sqrt{2}}\bigg[1 \pm  \prod_{p\alpha\neq k\sigma} (1-2\hat n_{p\alpha}) \bigg](1-\hat n_{k\sigma}),\nonumber\\
 \hat n_{k\sigma} (1-\hat n_{q\beta})\hat P^\pm &=  \frac{1}{\sqrt{2}}\bigg[1 \mp  \prod_{p\alpha\neq k\sigma,\,q\beta} (1-2\hat n_{p\alpha}) \bigg]\hat n_{k\sigma}(1-\hat n_{q\beta}), \nonumber \\
 \hat n_{k\sigma} \hat n_{q\beta}\hat P^\pm &=  \frac{1}{\sqrt{2}}\bigg[1 \pm  \prod_{p\alpha\neq k\sigma,\,q\beta} (1-2\hat n_{p\alpha}) \bigg]\hat n_{k\sigma}\hat n_{q\beta}, \nonumber
\end{align}}
\noindent
one finds
\begin{align}
 \langle \hat n_{k\sigma} \rangle_{even/odd} &= F_{k\sigma} \frac{1 \mp  \prod_{p\alpha\neq k\sigma} (1-2 F_{p\alpha})}{1 \pm  \prod_{p\alpha} (1-2 F_{p\alpha})}.
\end{align}
In the important limit of weak excitation, $F_{k\sigma} \ll 1 $, we can approximate this expression as follows
\begin{align}
 \langle \hat n_{k\sigma} \rangle_{even/odd} &= F_{k\sigma} \frac{1\mp \exp(-2\sum_{p\alpha \neq k \sigma} F_{p\alpha})}{1\pm \exp(-2\sum_{p\alpha} F_{p\alpha})}.
\end{align}
As mentioned in the main text, we assume that 
the level splitting in the quantum dot is smaller than temperature and bias voltage. Under these conditions many quasiparticle states
are always occupied, and we may further approximate 
$\sum_{p\alpha \neq k \sigma} F_{p\alpha} \approx \sum_{p\alpha}F_{p\alpha}=N_{qp}$. 
Hence we obtain
\begin{align}
  \label{eq_appendix::qp_occupation_even_odd_1}
\langle \hat n_{k\sigma} \rangle_{n} &= F_{k\sigma} [\tanh(N_{qp})]^{(-1)^n}\equiv \mathcal{A}_nF_{k\sigma}.
\end{align}
In order to find out under which conditions the parity effect becomes important, we plot  
the quasiparticle numbers in the even and odd states, 
\begin{eqnarray}
N_{qp}^n=\sum_{k\sigma} \langle \hat n_{k\sigma} \rangle_{n}={\cal A}_n N_{qp},
\label{Nqp_An}
\end{eqnarray}
versus $N_{qp}$ in Fig.~\ref{fig::appendix_Nqp}. 
It is clear for this plot that the parity effect needs 
to be taken into account for $N_{qp} \lesssim 2$. 

Subsequently we will also need the following expectation values
\begin{align}
\nonumber
 &\langle \hat n_{k\sigma} \hat n_{q\beta} \rangle_{even/odd} = F_{k\sigma}  F_{q\beta}\frac{1\pm \exp(-2\sum_{p\alpha \neq k \sigma,q\beta} F_{p\alpha})}{1\pm \exp(-2\sum_{p\alpha} F_{p\alpha})}, \\
 &\langle \hat n_{k\sigma} (1-\hat n_{q\beta}) \rangle_{even/odd} \nonumber\\&= F_{k\sigma}  (1-F_{q\beta})   \frac{1\mp \exp(-2\sum_{p\alpha \neq k \sigma,q\beta} F_{p\alpha})}{1\pm \exp(-2\sum_{p\alpha} F_{p\alpha})}. 
\label{eq_appendix::qp_double_occupation}
\end{align}
Employing the same set of approximations to this combination, we arrive at simple results
$\langle \hat n_{k\sigma} \hat n_{q\beta} \rangle \approx F_{k\sigma}  F_{q\beta}$ and $\langle \hat n_{k\sigma} (1-\hat n_{q\beta}) \rangle_n \approx \mathcal{A}_nF_{k\sigma}  (1-F_{q\beta})$.

\section{Tunneling rates and master equation}
\label{appendix::tunneling_rates}

In this appendix we will derive the master equation for the occupation probabilities
\begin{eqnarray}
p_n={\rm Tr}(\hat \rho^n),\;\;{\rm where}\;\;
\hat \rho^n=\hat \rho_{qp}^n\hat \rho_L\hat\rho_R\hat \rho_p,
\end{eqnarray}
of the charging states of the quantum dot. Let $\hat P_n=|n\rangle \langle n|$ be the projector onto charging state $|n\rangle$, then $ p_n={\rm Tr}(\hat P_n\hat \rho)$.
In order to keep track of the charge and the quasiparticle excitations on the dot we decompose the tunneling Hamiltonian into terms  
\begin{align}
 H^{++}_{k\sigma}&=\sum_{rk'} (t_{kk'}^r)^* u_k\, \hat T^\dagger e^{i\phi} \gamma_{k\sigma}^\dagger c_{rk'\sigma} \\
 H^{+-}_{k\sigma}&=\sum_{rk'} (t_{kk'}^r)^* \sigma \, v_k\, \hat T^\dagger e^{i\phi} \gamma_{-k-\sigma} c_{rk'\sigma},
\end{align}
and $H^{\alpha\beta}_{k\sigma}=(H^{-\alpha,-\beta}_{k\sigma})^\dagger$, with $\alpha,\,\beta=\pm$, such that $H_T= \sum_{k\sigma}\sum_{\alpha\beta}H^{\alpha\beta}_{k\sigma}$. Here $H^{+-}_{k\sigma}$ for example adds a charge and removes an excitation $\{-k,-\sigma\}$ on the dot. With Eq.~\eqref{eq_appendix::liouville} we obtain 
\begin{align}
\label{eq_appendix::master_eq_simple}
 \frac{d}{dt} p_n &= 2{\rm Re} \int_{-\infty}^t dt'\sum_{\alpha \beta}\sum_{k\sigma}\nonumber \\ &\times {\rm Tr}\big( H^{\alpha\beta}_{k\sigma}(t)H^{\bar\alpha\bar\beta}_{k\sigma}(t')\hat \rho_{n+\alpha}(t) 
		    - H^{\bar\alpha\bar\beta}_{k\sigma}(t')H^{\alpha\beta}_{k\sigma}(t)\hat \rho_{n}(t) \big)
\end{align}
where $\bar\alpha=-\alpha$. The contractions can be readily calculated using \eqref{eq_appendix::qp_occupation_even_odd_1}. As an example we choose the combination 
\begin{align}
I_1&\equiv 2{\rm Re} \sum_{k\sigma}\int_{-\infty}^t dt' {\rm Tr} \big(H^{++}_{k\sigma}(t)H^{--}_{k\sigma}(t')\hat \rho_{n+1}(t)\big)  \nonumber \\
&=\sum_{rkk'\sigma} |t_{kk'}^r|^2 u_k^2(1-f_{rk'\sigma})  \mathcal{A}_{n+1} F_{k\sigma} p_{n+1}\nonumber \\ & \times 2{\rm Re}  \int_{-\infty}^t dt' e^{-i(\xi_{rk'\sigma}+eV_r-eV_\phi(t)-E_{n+1}+E_n-E_{k})(t-t')}.
\end{align}
Here we introduced the charging energy $E_n=E_C(n-n_g^0)^2$, the distribution function of the leads, $f_{rk'\sigma}$, and linearized the time dependence of $\phi$, i.e. $\phi(t)-\phi(t')\approx eV_\phi(t) (t-t')$. 
In order to perform the time-integral we use ${\rm Re}\int_{-\infty}^0 d\tau e^{ix\tau+\eta\tau}=\eta/(x^2+\eta^2)\rightarrow \pi\delta(x)\,(\eta\rightarrow 0)$. Converting the $k'$-sum into an integral, i.e. $\sum_{k'}\rightarrow \mathcal{N}_F V\int d\xi'$, yields
\begin{align}
\label{eq_appendix::example1}
I_1&=  2\pi \mathcal{N}_FV \sum_{rk\sigma} |t^r|^2 p_{n+1}\nonumber \\
 &\times \left[u_{k}^2 \mathcal{A}_n F_{k\sigma} [1-f(-eV_r+eV_\phi(t)+E_{n+1} -E_{n}+ E_k)] \right. . 
\end{align}
In this way all various combinations are calculated. They are simmetrized by the transition rates $W_{n\mp 1,n}=\sum_r W^r_{n\mp 1,n}$ for transitions from charging states $n$ to $n\mp 1$,  
\begin{align}
  \label{eq_appendix::rates}
 &W_{n-1,n}^r (t)=\sum_{k\sigma} \frac{1}{e^2R_T^r\mathcal{N}_FV} \\
  &\times \big\{u_{k}^2 \mathcal{A}_n F_{k\sigma} [1-f(-eV_r-E_{n-1}+E_n+eV_\phi(t) + E_k)]  \nonumber \\   &+ v_k^2 \left(1-\mathcal{A}_n F_{k\sigma}\right) \nonumber \\ &\times[1-f(-eV_r-E_{n-1}+E_n+eV_\phi(t) - E_k)]\big\}, \nonumber \\ 
  \label{eq_appendix::rates_2}
&W_{n+1,n}^r (t)=\sum_{k\sigma}  \frac{1}{e^2R_T^r\mathcal{N}_FV} \\
  &\times \big\{u_{k}^2 \left(1-\mathcal{A}_n F_{k\sigma}\right) f(-eV_r-E_{n}+E_{n+1}+eV_\phi(t) + E_k)  \nonumber \\   &+  v_k^2 \mathcal{A}_n F_{k\sigma}f(-eV_r-E_{n}+E_{n+1}+eV_\phi(t) - E_k)\big\}. \nonumber  
\end{align}

\noindent 
Here we introduced the tunneling resistances $R_T^r$ as follows: $(e^2R_T^r)^{-1}=2\pi(\mathcal{N}_FV)^2|t_r|^2$. After all these transformations Eq.~\eqref{eq_appendix::master_eq_simple} acquires the form \eqref{eq::master_pn} given in the main text.

\section{Tunneling current}
\label{appendix::tunneling_current}

The current through lead $r$ is given by the expectation value $I_r=e\big\langle\frac{d}{dt}\sum_{k\sigma} c^\dagger_{rk\sigma} c_{rk\sigma}\big\rangle$ with the electron operators $c_{rk\sigma}$ and $c^\dagger_{rk\sigma}$ of lead $r=L,R$. Observing that the following commutator relation applies, $[\sum_{k\sigma} c^\dagger_{rk\sigma} c_{rk\sigma},\sum_{p,\alpha\beta} H^{\alpha\beta}_{p\sigma}]=-\sum'_{p,\alpha\beta} \alpha H^{\alpha\beta}_{p\sigma}$, one obtains{ within} the Born-Markov approximation
\begin{align}
  \label{eq_appendix::current}
 I_r (t)&= -e\int_{-\infty}^t dt'\sum'_{k\sigma,\alpha\beta}\, {\rm Tr}( \alpha H_{k\sigma}^{\alpha\beta}(t)[H_{k\sigma}^{\bar\alpha\bar\beta}(t'),\hat \rho(t)]), \nonumber \\
 &= -2e{\rm Re}\int_{-\infty}^t dt'\sum'_{k\sigma,\alpha\beta}\, {\rm Tr}( \alpha H_{k\sigma}^{\alpha\beta}(t)H_{k\sigma}^{\bar\alpha\bar\beta}(t')\hat \rho(t)).
\end{align}
Here the prime in the sums shall indicate that exclusively tunneling events from and to lead $r$ are considered. 
Eq.~\eqref{eq_appendix::current} is very similar to Eq.~\eqref{eq_appendix::master_eq_simple} and therefore the current can be expressed via the transition rates \eqref{eq_appendix::rates} and \eqref{eq_appendix::rates_2},
\begin{align}
 I_r (t)= e \sum_n [W_{n+1,n}^r(t) - W_{n-1,n}^r(t)]\,p_n(t).
\end{align}

\section{Kinetic equation}
\label{appendix::kinetic_equation}

In order to evaluate the kinetics of the quasiparticles one has ask for the probability to find $n$ electrons and a quasiparticle in the state $k\sigma$, i.e. $\langle \hat n_{k\sigma} \hat P_n \rangle$,
\begin{align}
 \label{eq_appendix::kinetic_equation}
 \frac{d}{dt} \big(\hat n_{k\sigma} \hat P_n\big) \bigg|_{tun}&= i\big([H_I,\hat n_{k\sigma}]\hat P_n + \hat n_{k\sigma}[H_I,\hat P_n] \big)
\end{align}

\noindent
In this case both the tunneling and the electron-phonon interaction have to be considered, $H_I=H_T+H_{ep}$. 

\subsection{Tunneling}
First we consider the contribution coming from the tunnel Hamiltonian,
\begin{align}
 \label{eq_appendix::kinetic_equation_tun}
 \frac{d}{dt} \big(\hat n_{k\sigma} \hat P_n\big) \bigg|_{tun}&= i\big([H_T,\hat n_{k\sigma}]\hat P_n + \hat n_{k\sigma}[H_T,\hat P_n] \big)
\end{align}
By summing up all charging states $n$ the second term in Eq.~\eqref{eq_appendix::kinetic_equation_tun} vanishes due to charge conservation. The first commutator gives 
\begin{align}
 \frac{d}{dt} \sum_n \big\langle\hat n_{k\sigma} \hat P_n\big\rangle \bigg|_{tun} &=  -2{\rm Re}\int_{-\infty}^t dt' \sum_n \sum_{\alpha\beta} \nonumber \\ &\times {\rm Tr} \big(\beta H_{k\sigma}^{\alpha\beta}(t)H_{k\sigma}^{\bar\alpha\bar\beta}(t')\hat \rho_{n}(t)\big),
\end{align}
which is again very similar to \eqref{eq_appendix::master_eq_simple} and \eqref{eq_appendix::current}. Therefore, without going into detail, we get
{
\begin{align}
\label{eq_appendix::qp_tunneling}
 &\frac{d}{dt}\sum_n\big\langle\hat n_{k\sigma} \hat P_n\big\rangle \bigg|_{tun} = 
\sum_{nr} \frac{p_n}{e^2R_T^r\mathcal{N}_FV} \\
&\times \left\{  -\left[1-f(-eV_r - E_{n-1} + E_n + eV_\phi(t) + E_{k})\right]u^2_{{k\sigma}}\mathcal{A}_nF_{{k\sigma}} \right. \nonumber\\
& +  \left[1-f(-eV_r - E_{n-1} + E_n + eV_\phi(t) - E_{{k}})\right] v^2_{{k\sigma}}(1-\mathcal{A}_nF_{{k\sigma}})\nonumber\\
&+  f(-eV_r + E_{n+1} - E_n - eV_\phi(t) + E_{{k}})u^2_{{k\sigma}}(1-\mathcal{A}_nF_{{k\sigma}}) \nonumber\\
&- \left. f(-eV_r + E_{n+1} - E_n - eV_\phi(t) - E_{{k}})v^2_{{k\sigma}} \mathcal{A}_nF_{{k\sigma}}\right\}.\nonumber
\end{align}
}

\subsection{Inelastic phonon scattering}
In order to derive the electron-phonon collision integral, we repeat the same analysis as in the previous subsection replacing the tunnel Hamiltonian in Eq.~\eqref{eq_appendix::kinetic_equation} by the electron-phonon interaction \eqref{Hep}.
We start by decomposing the electron-phonon Hamiltonian into parts, 
\begin{align}
 H_{ep} &= \sum_{qk\sigma} h_{k+q,k}^\sigma + h.c. \\
 h_{k+q,k}^\sigma&= \big(S_{k+q, k}^{\sigma}+R_{k+q, k}^{\sigma}\big) \hat \varphi_q\\
 S_{k+q, k}^{\sigma} &=  g_{k+q,k}(u_{k+q}u_{k}-v_{k+q}v_k) \gamma^\dagger_{k+q,\sigma}\gamma_{k\sigma} \\
 R_{k+q, k}^{\sigma} &=  g_{k+q,k}(u_{k+q}v_{k}+v_{k+q}u_k)  \gamma_{k+q,\sigma}^\dagger\overline \gamma_{k\sigma}^\dagger \\
 \hat \varphi_q &= b_q+b_{-q}^\dagger . 
\end{align}
Mind that electron-phonon interaction does not change the charge on the dot. Thus we only have to consider the commutator
$
 [\hat n_{k\sigma},H_{ep}]=\sum_{q} h^\sigma_{k,k-q} - h.c.
$,
\begin{align}
 \frac{d}{dt} \sum_n\big\langle\hat n_{k\sigma} \hat P_n\big\rangle \bigg|_{ep}
 &= -2 {\rm Re}  \int_{-\infty}^t dt' \sum_{nq} \nonumber \\ &\times {\rm Tr} \bigg(h_{k,k-q}^\sigma(t)[h_{k,k-q}^\sigma(t')]^\dagger \hat \rho_n(t) \nonumber \\ &- [h_{k,k-q}^\sigma(t)]^\dagger h_{k,k-q}^\sigma(t') \hat \rho_n(t)   \bigg).
\end{align}
This equation involves contributions accounting for pair-braking/recombination and scattering. 
The contractions lead to the common collision integrals for electron-phonon interaction. For instance we obtain

\begin{align}
\label{eq_appendix::I2}
 I_2 &\equiv -2 {\rm Re}  \int_{-\infty}^t dt' \sum_{nq} \nonumber \\ &\times {\rm Tr} \big(S_{k,k-q}^\sigma(t)[S_{k,k-q}^\sigma(t')]^\dagger\hat \varphi_q(t)\hat\varphi_q^\dagger(t') \hat \rho_n(t) \big) \\
 &= -2{\rm Re} \int_{-\infty}^t dt' \sum_{nk'q} |g_{kk'} (u_{k}u_{k'}-v_{k}v_{k'})|^2 {\rm Tr} \big(\hat \rho_n(t)  \nonumber \\
 &\times\big\{  e^{i(E_k-E_{k'}-\omega_q)(t-t')}\hat n_{k\sigma}(1-\hat n_{k',\sigma}) (1+\hat N_q) \delta_{k',k-q} \nonumber \\
 &+  e^{i(E_k-E_{k'}+\omega_q)(t-t')}\hat n_{k\sigma}(1-\hat n_{k',\sigma}) \hat N_{q} \delta_{k',k+q} \big \}\big),\nonumber \\
\end{align}
with $\hat N_q = b^\dagger_q b_q$.
The last two lines correspond to processes where a quasiparticle in the state $k\sigma$ is scattered into the state $k'\sigma$ by emitting or absorbing a phonon. Performing the time-integral entails the energy conservation for each of these processes. 
On the other hand 
\begin{align}
\label{eq_appendix::I3}
 I_3 &\equiv -2 {\rm Re}  \int_{-\infty}^t dt' \sum_{nq} \nonumber \\ &\times {\rm Tr} \big(R_{k,k-q}^\sigma(t)[R_{k,k-q}^\sigma(t')]^\dagger\hat \varphi_q(t)\hat\varphi_q^\dagger(t') \hat \rho_n(t) \big) \\
 &= -2{\rm Re} \int_{-\infty}^t dt' \sum_{nk'q} |g_{kk'} (u_{k}v_{k'}+v_{k}u_{k'})|^2 {\rm Tr} \big( \hat \rho_n(t) \nonumber \\
 &\times\big\{  e^{i(E_k+E_{k'}-\omega_q)(t-t')}\hat n_{k\sigma}\hat n_{k'\bar \sigma} (1+\hat N_q) \delta_{k',-k+q}\}\big)\nonumber \\
\end{align}
corresponds to the recombination of two quasiparticles and the emission of a phonon with energy $\omega_q>2\Delta$.
With \eqref{eq_appendix::qp_double_occupation} we find
\begin{align}
  I_2 &= \frac{\pi b}{{\mathcal N_F}V} \sum_n\sum_{k'}   (E_k-E_{k'})^2 \bigg(1+\frac{\xi_k\xi_{k'}}{E_kE_{k'}}-\frac{\Delta^2}{E_kE_{k'}} \bigg) \nonumber \\ &\times \mathcal A_n F_{k\sigma}(1-F_{k',\sigma}) (1+n^B_{E_k-E_{k'}}){\rm sign}(E_k-E_{k'}) p_n ,\\
  I_3 &= -\frac{\pi b}{{\mathcal N_F}V} \sum_n\sum_{k'}   (E_k+E_{k'})^2 \bigg(1-\frac{\xi_k\xi_{k'}}{E_kE_{k'}}+\frac{\Delta^2}{E_kE_{k'}} \bigg) \nonumber \\ &\times F_{k\sigma}F_{k'\bar\sigma} (1+n^B_{E_k+E_{k'}}) p_n .
\end{align}
Here $n^B_\omega=1/[\exp(\beta\omega)-1]$ is the equilibrium Bose-distribution. 
We further assume that the Fermi-surface averaged electron-phonon coupling matrix is absorbed in a single constant $b$, so that
\begin{align}
\label{def_b}
 \frac{1}{\mathcal{N}_FV}\sum_{pp'} |g_{p,p'}|^2 \delta(\xi_p)\delta(\xi_{p'})\delta(\omega-\omega_{p-p'})\approx b \omega^2 \theta(\omega).
\end{align}
Here we essentially assumed an isotropic electron-phonon coupling $g_{p,p'}$ and a Debye phonon density of states typical for acoustic phonons. Finally we obtain
{\small
\begin{align}
\label{eq_appendix::qp_ep}
 &\frac{d}{dt} \sum_n\big\langle\hat n_{k\sigma} \hat P_n\big\rangle \bigg|_{ep} =  \frac{\pi b}{{} \mathcal{N}_FV} \sum_{np}\\
  &\,\,\,\times (E_{k}+E_{p})^2 \left[1-\frac{\xi_{k}\xi_{p}}{E_{k}E_{p}}+\frac{\Delta^2}{E_{k}E_{p}}\right] \nonumber \\
				 &\times \left[(1-F_{k\sigma})(1-F_{p\bar\sigma})n^B_{E_{k}+E_{p}} -F_{k\sigma}F_{p\bar \sigma}(1+n^B_{E_{k}+E_{p}}) \right]p_n \nonumber  \\
	       &+ (E_{p}-E_{k})^2{\rm sign}(E_{p}-E_{k})\left[1+\frac{\xi_{k}\xi_{p}}{E_{k}E_{p}}-\frac{\Delta^2}{E_{k}E_{p}}\right]\nonumber \\
				 &\times  \left[F_{p\sigma}(1-F_{k\sigma})(1+n^B_{E_{p}-E_{k}})-F_{k\sigma}(1-F_{p\sigma})n^B_{E_{p}-E_{k}} \right]\mathcal{A}_np_n . \nonumber
\end{align}
 }
\noindent
In the limit of many excitations and $\mathcal{A}_n\rightarrow 1$ the Eq.~\eqref{eq_appendix::qp_ep} reduces to the familiar form \cite{Baryakhtar_1979}.  

\section{Rothwarf-Taylor equation}

In this section we are going to derive a simple kinetic equation for the quasiparticle number $N_{qp}$ in the same way as 
Rothwarf and Taylor did in Ref.~\cite{Rothwarf_Taylor_1967}. 
For simplicity we neglect the charge imbalance and assume that $F_{k\sigma}=F_{k'\sigma}$ if $\xi_k=-\xi_{k'}$. 
We integrate the kinetic equation (\ref{eq::full_boltzmann}) over the quasiparticle energies.
First we observe that those contributions in Eqs.~\eqref{eq_appendix::qp_tunneling} and \eqref{eq_appendix::qp_ep}, 
which do not depend on the quasiparticle distribution function, can be combined in the injection rate 
\begin{align}
 I_n^{qp} &= \sum_{k s=\pm 1} \frac{1}{e^2R_T^r\mathcal{N}_F\mathcal{V}} \\
&\left[1-f(-eV_r - E_{n-1} + E_n + eV_\phi(t) -E_k) \right. \nonumber\\
&+\left. f(-eV_r + E_{n+1} - E_n + eV_\phi(t) +E_k) \right]. \nonumber
\end{align}
Next we define the tunneling rate $ \Gamma^r_n$, which effectively describes the relaxation
of the quasiparticle distribution function via the tunneling in or out of the leads.
We assume that all excited quasiparticles have the energies just above the superconducting gap $\Delta$.
This assumption is justified by the fact that 
in our setup quasiparticles are injected close to the gap and the electron-phonon relaxation is sufficiently strong. 
Keeping that in mind we make the following approximation:
\begin{align}
 \Gamma^{qp}_n &= \sum_{k\sigma s=\pm 1} \frac{1}{2e^2R_T^r\mathcal{N}_F\mathcal{V}} \\
&\left[1-f(-eV_r - E_{n-1} + E_n + eV_\phi(t) + sE_k) \right. \nonumber\\
&+\left. f(-eV_r + E_{n+1} - E_n + eV_\phi(t) + sE_k) \right]\frac{\mathcal{A}_n F_{k\sigma}}{N_{qp}\mathcal{A}_n} \nonumber  \\
 &\approx \sum_{s=\pm 1} \frac{1}{2e^2R_T^r\mathcal{N}_FV} \nonumber\\
&\left[1-f(-eV_r - E_{n-1} + E_n + eV_\phi(t) + s\Delta) \right. \nonumber\\
&+\left. f(-eV_r + E_{n+1} - E_n + eV_\phi(t) + s\Delta) \right]. \nonumber
\end{align}
Considering now the electron-phonon collision integral, we note that those terms which conserve the number of quasiparticles vanish upon the integration. 
Next, at sufficiently low temperatures, and also since the phonon bath stays in equilibrium, the  pair-braking processes are suppressed. 
Thus only recombination contributes to the quasiparticle relaxation. The rate of recombination reads 
\begin{align}
 \kappa &=  \frac{\pi b}{{} \mathcal{N}_F\mathcal{V}} \sum_{k\sigma p} (E_{k}+E_{p})^2 \left[1+\frac{\Delta^2}{E_{k}E_{p}}\right] \frac{F_{k\sigma}}{N_{qp}}\frac{F_{p\bar \sigma}}{N_{qp}}\nonumber \\
 &\approx  \frac{4\pi b \Delta^2}{{} \mathcal{N}_F\mathcal{V}}.
\end{align}

Combining all these results we arrive at the Rothwarf-Taylor equation
\begin{align}
 \frac{d}{dt}\left[\sum_n p_n N_{qp} \mathcal{A}_n \right]&= \sum_n p_n\,\left[I_n^{qp} - \Gamma_n^{qp} N_{qp}\mathcal{A}_n - \kappa N_{qp}^2 \right].
\end{align}

\section{Relation to the formalism of Ref. \cite{Maisi_2012}}
\label{QP_rates}

In this appendix we demonstrate the equivalence of the approach used in this paper to that of Ref. \cite{Maisi_2012} by 
showing that Eq. (\ref{eq::Rothwarf_Taylor}) also follows from the latter.
In Ref. \cite{Maisi_2012} the system dynamics is described in terms of joint probability distribution
of electron number, $n$, and quasiparticle number, $m$, which we denote as
$p_{nm}$. It satisfies the master equation
\begin{eqnarray}
\frac{dp_{nm}}{dt} &=& 
W_{n,n-1}^{m,m-1} p_{n-1,m-1} + W_{n,n-1}^{m,m+1} p_{n-1,m+1} 
\nonumber\\ &&
+\, W_{n,n+1}^{m,m+1}p_{n+1,m+1} + W_{n,n+1}^{m,m-1}p_{n+1,m-1}
\nonumber\\ &&
-\, \big(W_{n+1,n}^{m+1,m} + W_{n+1,n}^{m-1,m} 
\nonumber\\ &&
+\, W_{n-1,n}^{m-1,m}+ W_{n-1,n}^{m+1,m}\big)p_{nm}.
\label{Pnm} 
\end{eqnarray} 
For the sake of simplicity  here we ignore the electron-phonon interaction. 
In this equation the rate $W_{n+1,n}^{m+1,m}$, for example, describes the tunneling of one electron
into a superconducting island with simultaneous creation of a quasiparticle, while
the rate $W_{n+1,n}^{m-1,m}$ describes the electron tunneling into the island
accompanied by an annihilation of a quasiparticle.  These two rates are
defined as follows 
\begin{eqnarray}
W_{n+1,n}^{m\pm 1,m}= W_{n+1,n}^{m\pm 1,m; L}+W_{n+1,n}^{m\pm 1,m; R},
\end{eqnarray} 
\begin{eqnarray}
 W_{n+1,n}^{m+1,m;r} = \sum_{\sigma} \int d\xi\, w_{n+1,n}^{r}(E)\frac{1-F_{\xi\sigma}^{(m)}}{2}\left(1+\frac{\xi}{E}\right),
 \nonumber
\end{eqnarray}
\begin{eqnarray}
 W_{n+1,n}^{m-1,m;r} = \sum_{\sigma} \int d\xi\,  w_{n+1,n}^{r}(-E)\frac{F_{\xi\sigma}^{(m)}}{2}\left(1-\frac{\xi}{E}\right).
\nonumber
\end{eqnarray}
Here the distribution function $F_{\xi\sigma}^{(m)}$ is normalized in such a way that
\begin{eqnarray}
{\cal N}_F{\cal V}\sum_\sigma\int d\xi\,F_{\xi\sigma}^{(m)}=m.
\end{eqnarray}
The remaining rates are defined similarly. 

The specific form of the distribution function $F_{\xi\sigma}^{(m)}$ is not important
as long as $F_{\xi\sigma}^{(m)}\ll 1$. Indeed in this limit and in the absence
of charge imbalance we may approximate the rates in the following way
\begin{eqnarray}
 W_{n+1,n}^{m+1,m;r} = \int d\xi\, w_{n+1,n}^{r}(E)
- m\,\frac{w_{n+1,n}^{r}(\Delta)}{2{\cal N}_F{\cal V}},
\label{W11}
\end{eqnarray}
\begin{eqnarray}
 W_{n+1,n}^{m-1,m;r} = m\,\frac{w_{n+1,n}^{r}(-\Delta)}{2{\cal N}_F{\cal V}},
\label{W1-1}
\end{eqnarray}
and similarly for all remaining rates.

Next, we multiply Eq. (\ref{Pnm}) by the quasiparticle number $m$ and perform the
summation over both $m$ and $n$. After some manipulations we arrive at the result
\begin{eqnarray}
\frac{d}{dt}\sum_{mn}\big[ m p_{nm} \big] = \sum_{mn} p_{nm} 
\big[ W^{m+1,m}_{n+1,n} - W^{m-1,m}_{n+1,n} 
\nonumber\\
-\, W^{m-1,m}_{n-1,n} + W^{m+1,m}_{n-1,n} \big].
\label{N_qp_n}
\end{eqnarray}

Next, we introduce the occupation probability of a state with $n$ electrons trapped in the island,
$p_n=\sum_m p_{nm}$, and the average number of excited quasiparticles in this state,
$N_{qp}^{n}=\sum_{m} m p_{nm}/p_n$. The latter parameter should be equalized with the number of
quasiparticles defined in the Eq. (\ref{Nqp_An}), i.e. we put $N_{qp}^{n} = {\cal A}_nN_{qp}$.
Combining Eqs. (\ref{W11},\ref{W1-1}) and (\ref{N_qp_n}) we arrive at the result 
\begin{eqnarray}
\frac{d}{dt}\left[N_{qp} \sum_n p_n{\cal A}_n  \right] = \sum_n p_n\big[ I_n^{qp} - \Gamma_n^{qp} N_{qp} {\cal A}_n \big],
\label{Nqp_11}
\end{eqnarray}
The Eq. (\ref{Nqp_11}) coincides with the Eq. (\ref{eq::Rothwarf_Taylor}) of the main text
with omitted recombination term.
Thus we have indeed demonstrated the equivalence of the two approaches.

\end{document}